\documentstyle[prb, aps, amssymb, epsfig, pifont]{revtex}
\begin{document}
\draft


\wideabs{
\title{Glassy Slowing of Stripe Modulation 
in (La,Eu,Nd)$_{2-x}$(Sr,Ba)$_{x}$CuO$_{4}$: ~~~~~~~~~~~~~~~~~~~~~~~~~
A $^{63}$Cu and
$^{139}$La NQR  Study Down to 350\,mK}
\author{A.W. Hunt, P.M. Singer, A.F. Cederstr\"om, and T. Imai}
\address{Department of Physics and Center for Materials Science and 
Engineering, M.I.T., Cambridge, MA 02139}
\date{\today}
\maketitle


\begin{abstract}
$^{63}$Cu and $^{139}$La nuclear quadrupole resonance and Zeeman
perturbed nuclear magnetic resonance experiments are performed on the
striped phase of the high temperature superconductors 
La$_{2-x}$Ba$_{x}$CuO$_{4}$ and
La$_{2-x-y}$(Nd,Eu)$_{y}$Sr$_{x}$CuO$_{4}$.  The first goal of the
present study is to utilize the fact that ordered Cu magnetic moments
exert a static hyperfine field on the
$^{63}$Cu and $^{139}$La nucleii to deduce the charge density
and ordered moment within the CuO$_2$ planes.
A hyperfine broadened NQR lineshape is observed in both
La$_{2-x}$Ba$_{x}$CuO$_{4}$ and 
La$_{1.80-x}$Eu$_{0.20}$Sr$_{x}$CuO$_{4}$ for $x\approx\frac{1}{8}$.
Detailed numerical analysis of the $^{63}$Cu NQR lineshape
establishes  that widely accepted models of periodic sinusoidal or
square-well shaped  modulations of spin density waves with maximum moment
$\sim$\,0.3$\mu_{B}$,  as inferred from elastic neutron scattering
and $\mu$SR measurements, can {\it not} account for the NQR lineshape 
unless we assume a relatively small ordered moment $\sim$\,0.15$\mu_{B}$ 
with a comparably large distribution.
The second goal of the present work is to establish the temperature 
dependence of the fluctuation frequency scale of stripes. We find that
the the fraction of missing $^{63}$Cu NQR intensity below charge
ordering temperature
$T_{charge}$ accurately tracks the temperature dependence of the  charge
order parameter as
measured by scattering methods.  By fitting a {\it single} model to the
temperature dependences of the wipeout fraction $F(T)$ for
$^{63}$Cu and
$^{139}$La NQR, the spin order  parameter measured by elastic neutron
scattering, and the $\mu$SR data, we deduce the spatial distribution of
the spin fluctuation frequency scale $\Gamma$ and its temperature
dependence.  These results indicate  that as soon as charge dynamics slow
down, the spin fluctuations begin to slow dramatically with spin
stiffness $2\pi\rho_s^{eff}\sim200$\,K.  By
extending the analysis to other hole concentrations, we demonstrate
a qualitative difference in the spatial variation of electronic states
induced by slowing charge dynamics above and below $x=\frac{1}{8}$. 
\end{abstract}

\pacs{76.60, 74.25-N, 74.72}
}

\section {Introduction}
\label{sec_introduction}
The electronic phase diagram of high $T_{c}$ cuprates 
involves a number of different phases and crossovers, 
including an undoped antiferromagnetic phase (N\'eel state), 
an insulator-metal crossover, a spin-glass phase, a pseudo-gap phase, and 
an over-doped metallic regime, in  addition to the superconducting phase.
The complexity of the phase  diagram makes it  difficult to identify the
superconducting mechanism.  Some of the  aformentioned phases and
crossovers may be irrelevant to the superconducting  mechanism, while
others may prove to be critical.\cite{theory}  Further experimental efforts
are necessary to elucidate the  properties of each phase, as  well as the
nature of transitions and crossovers  between adjacent phases.  

A relatively new addition to the already complex electronic phase 
diagram of high $T_c$ cuprates is the so-called {\it stripe phase} near 
the magic hole concentration  $x\approx\frac{1}{8}$.  It has been
demonstrated by elastic neutron  scattering experiments on Nd co-doped
La$_{2-x}$Sr$_x$CuO$_4$  by Tranquada, Sternlieb, Axe, Nakamura, and
Uchida, that doped holes microscopically phase segregate into hole
rivers forming a charge density wave (CDW).\cite{tranquada1} The hole
rivers become antiphase boundaries between hole-poor segments in which Cu
spins form a short range spin density wave (SDW) order. 
Recent measurements based on neutron scattering,
\cite{suzuki1998,kimura,wakimoto,lee,tranquada2,tranquada3,tranquada4}
NQR,\cite{hunt,singer,curro,teitelbaum,julien} ESR,\cite{rameev,kataev}
x-ray scattering,\cite{zimmermann,niemoller} charge
transport,\cite{ando-boebinger,boebinger,noda,ichikawa,uchida,ichikawa-thesis}
photo emission spectroscopy\cite{Fujimori,zhou} and
$\mu$SR\cite{luke,kumagai,nachumi,niedermayer,kojima,savici}  detect
experimental signatures of the stripe  instabilites in the  underdoped
regime.

A major difficulty in applying NMR techniques 
to investigate the stripe phase of high $T_{c}$ cuprates is in 
the {\it glassy} nature of the slowing down of 
stripes that is observed for most striped cuprates 
\cite{tranquada4,nachumi} (one exception being 
La$_{2}$CuO$_{4+\delta}$ \cite{lee,savici}).  Unlike ordinary 
second order phase transitions, the slowing of stripe fluctuations 
is very gradual and inhomogeneous.  
This means that the experimental signature of stripe freezing 
appears at different temperatures depending on the characteristic 
frequency of the experimental probe.  If one uses a probe with a lower 
frequency scale, the stripe anomaly takes place at a lower temperature.
Furthermore, the glassy freezing of the stripes do not show typical
signatures  of phase transitions observed by NMR, such as a divergence of
NMR  spin-lattice and spin-spin relaxation rates $1/T_{1}$ and $1/T_{2}$ 
at the on-set of freezing.  That 
explains why earlier NMR studies of underdoped high $T_{c}$ 
cuprates carried out before the discovery of stripes by Tranquada {\it
et\,al.}\ by neutron scattering failed to capture the various signatures
of stripe  instabilities in a convincing manner, even though there are 
some results that  seem, retrospectively,  consistent with stripe
anomalies.\cite{imai89,yoshimura92,tou,cho,chou,goto,ohsugi1,ohsugi2}  

Very recently, we demonstrated that slowing of stripes 
in La$_{2-x}$Sr$_x$CuO$_4$ with and without Nd or Eu rare earth co-doping 
and in  La$_{1.875}$Ba$_{0.125}$CuO$_4$ can be easily captured by
measuring the anomalous  reduction (i.e. {\it wipeout}) of the $^{63}$Cu
NQR intensity.\cite{hunt,singer}  

In our earlier publications,\cite{hunt,singer} we left some of the
important issues  unexplored.  First, if stripes keep slowing down with
decreasing  temperature, what kind of spin  and charge density modulation
does the stripe phase exhibit in the  ground state {\it at NMR/NQR time
scales}?  Second, what is the spatial distribution of the stripe
fluctuation frequency? Third, how does the hole concentration affect the
slowing of stripes?  In order to address  these issues, we now report a
more detailed $^{63}$Cu and $^{139}$La NQR study of 
La$_{1.875}$Ba$_{0.125}$CuO$_4$ and (Nd, Eu) co-doped
La$_{2-x}$Sr$_x$CuO$_4$ near the magic hole concentration
$x\approx\frac{1}{8}$ down to 350\,mK.  In section
\ref{section_zeeman_perturbed},  we report Zeeman preturbed $^{63}$Cu and
$^{139}$La NQR lineshapes including  detailed numerical simulations.   Our
preliminary lineshape measurements on $^{63}$Cu isotope enriched  samples
conducted  for our earlier publication\cite{hunt} did not reveal any
noticeable  structures down to 1.7\,K other than a single peak, in
agreement with the  earlier finding by Tou {\it et\,al.}\ reported for a
$^{63,65}$Cu  mixed isotope sample.\cite{tou}  This made us suspect that 
stripes are still fluctuating quickly enough relative to NQR
time scales to {\it motionally  narrow}\cite{abragam,slichter} the 
lineshapes at 1.7 K.  Accordingly, we developed a top-loading $^{3}$He 
NMR system to conduct NQR lineshape measurements in a broad frequency 
range down to 350\,mK in the hope that any stripe excitations 
with energy $\Delta E/k_{B}\sim$1 K or higher would be suppressed. 
However, our new results at  350\,mK  demonstrate that a {\it periodic}
spatial modulation of spin and charge density hardly exists at NQR time 
scales even at 350\,mK.  This is in remarkable contrast with  elastic
neutron, x-ray and $\mu$SR measurements  that sucessfully detect short
range spin and charge order {\it at time scales that are three  to seven
orders of magnitude faster than NQR measurements}.  Our NQR  results
suggest that the short range stripe order is highly disordered or  that
extremely slow dynamic fluctuations  persist even at 350\,mK,  or probably
both.  In section \ref{section_slowing}, we analyze the distribution of
$^{139}$La nuclear relaxation $^{139}1/T_1$ and demonstrate that below
charge ordering, glassy slowing of the Cu spins fluctuations sets in and
results in large spatial distribution of $^{139}1/T_1$.  We also analyze
the wipeout of $^{63}$Cu and $^{139}$La.  We employ the renormalized 
classical scaling in the non-linear-$\sigma$-model  to fit the entire set
of existing data with small spin stiffness
($2\pi\rho_s^{eff}\lesssim200$\,K) which is reduced due to slowed charge
dynamics.  This analysis accounts for the  temperature dependences of not
only the $^{63}$Cu NQR wipeout, but also the
$^{139}$La NQR wipeout, the order parameter of spin stripes measured by
elastic neutron  scattering, the $\mu$SR asymmetry, and the recovery of
both $^{63}$Cu and $^{139}$La NQR intensity, all within a single
framework.  From the analysis, we estimate the spatial distribution of
stripe  fluctuations as a function of temperature.  We find that below
$T_{charge}$  the glassy slowing down of stripes results in three
orders of magnitude distribution in fluctuation frequencies in agreement
with our analysis of $^{139}1/T_1$.  This is in contrast to the well known
neutron results in the normal metallic state of
La$_{2-x}$Sr$_{x}$CuO$_{4}$, where the spin fluctuation energy scale
$h\Gamma$ is well-defined by a single value on the order of
$k_{B}T$, and that $h\Gamma$ varies roughly in proportion  to $T$ as
originally discovered by Keimer {\it et\,al.}\cite{keimer} (called {\it
$\omega/T$-scaling},  \cite{keimer,sternlieb} or {\it quantum critical
behavior} \cite{aeppli}).   In  section V, we extend the analysis to
deduce a diagram of temperature versus stripe fluctuation frequency for
various hole concentrations. 

\section{Experimental}
\label{sec_sample_preparation}

All of the single phase, polycrystalline samples of
La$_{2-x-y}$(Eu,Nd)$_y$(Ba,Sr)$_x$CuO$_4$ used in this study were prepared
using conventional solid state procedures.
We mix predried La$_{2}$O$_{3}$ (99.99$\%$), 
Eu$_{2}$O$_{3}$ (99.99$\%$), Nd$_{2}$O$_{3}$ (99.99$\%$),
BaCO$_{3}$ (99.95$\%$), SrCO$_{3}$ (99.99$\%$), and CuO (99.995$\%$) with 
correct nominal compositions, and grind by hand with an agate mortar and
pestle until an intimate mixture is obtained.  To make complicated
lineshape analysis feasible, some samples were enriched with either
$^{63}$Cu or $^{65}$Cu isotopes.  This first grinding usually consumes
40-60 minutes for the 1-4\,g of powder that we prepare.  A prereaction is
carried out for 20\,h in a box furnace at $850^{\circ}$C followed by
repeated regrindings and sinterings (also 20\,h) at temperatures between
$950^{\circ}$C and $1000^{\circ}$C.  Finally the samples are pelletized
and high temperature annealed in flowing O$_2$ gas at $1100^{\circ}$C to
$1150^{\circ}$C for 24\,h to 48\,h before a slow and controlled cooling
cycle that includes low temperature annealing at $800^{\circ}$C (24\,h) and
$500^{\circ}$C (24\,h).  By using a large number of grindings
(typically 5-8), we acheive homogenous, high quality polycrystaline
samples. The powder x-ray spectra reveal that all samples are single
phased. The measured lattice parameters show systematic variation through
the sample families as a function of hole concentration, revealing the
room temperature orthorhombic to tetragonal structural transition, and
are well within the range of previously published data.\cite{breuer}
To further characterize the prepared samples, magnetization measurements
are performed with a SQUID magnetometer at a field of 10\,Oe.  All
superconducting samples show a clear diamagnetic response, and $T_c$
and volume fraction are consistent with previous
measurements.\cite{buchner94}  Nonsuperconducting samples show no
Meissner shielding down to 5\,K.

All nuclear resonance data are taken with a homemade, phase coherent pulsed NMR
spectrometer using the standard $\pi/2-\tau-\pi$ pulse sequence with a typical
pulse separation time $\tau=12$ $\mu$s.  For measurements at and
above 1.7\,K we use a commercial cryostat and measure temperature with
calibrated Cernox and carbon-glass resistors.  In order to conduct Zeeman
perturbed NQR measurements in zero applied field down to 350\,mK, we
developed a top-loading NMR system utilizing a commercial single-shot
$^3$He refrigerator.  In those cases,  temperature is measured via a
calibrated ruthenium oxide temperature sensor.

\section{Zeeman Perturbed NQR Lineshape}
\label{section_zeeman_perturbed}

\subsection{Principles of Zeeman perturbed NQR}
\label{subsec_zeeman_principles}

\begin{figure}[ht]
\begin{center}
\epsfig{figure=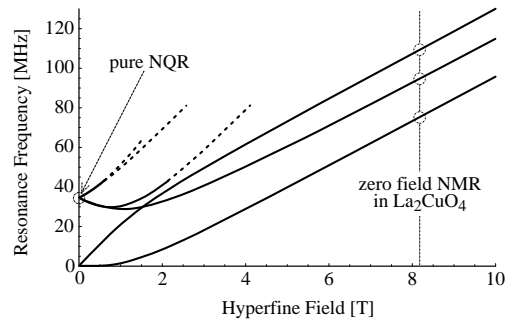,width=2.6in,angle=0}
\end{center}
\caption{Plot of $^{63}$Cu NQR ($I=\frac{3}{2}$) transition frequencies 
as a function of hyperfine field in the
ab-plane with an axially symmetric quadrupole tensor
$\nu_Q^{c}=34.5$\,MHz.  This chart is applicable for the case of $^{63}$Cu
resonance in La$_{1.875}$Ba$_{0.125}$CuO$_{4}$.  The dashed lines mark
transitions that occur with very low probability.  The intersects of
three solid curves and  the dotted vertical line at $H=8.2$\,T
correspond  to three zero field Cu NMR lines observed for
La$_{2}$CuO$_{4}$ by Tsuda  {\it et\,al.}\cite{tsuda}  (To be rigorous,
$\nu_Q^{c}=31.9$\,MHz at 1.3\,K in La$_{2}$CuO$_{4}$.)}
\label{fig_transition_energies}
\end{figure}

In the N\'eel ordered state of the undoped parent compounds of high
$T_{c}$ cuprates, Zeeman perturbed NQR has been sucessfully utilized to
measure the sublattice magnetization $\langle S_{i}\rangle$ and its
temperature dependence.  In antiferromagnets below the N\'eel temperature
$T_{N}$, the nuclear resonance frequency $f_{i}$ of the {\it i}-th site
depends on $\langle H_{i}\rangle$, the static component of hyperfine 
magnetic field at that  site.  Typically, $\langle H_{i}\rangle$
originates from the  hyperfine coupling $A^{ii}$ with the ordered
magnetic moment $\langle S_{i}\rangle$ at   the same site and from the
transferred hyperfine coupling $A^{ij}$ with the ordered magnetic moment
$\langle  S_{j}\rangle$ at the nearest neighbor sites.   
Since the hyperfine coupling constants can be determined based on NMR
measurements in the  paramagnetic state above $T_{N}$,  
one can utilize the nuclear resonance frequency $f_{i}$ to measure the 
hyperfine magnetic field $\langle H_{i}\rangle$, and hence the 
sublattice magnetization $\langle S_{i}\rangle$.\cite{yasuoka-old,benedek} 
This technique has been employed successfully in the N\'eel state of
several undoped parent compounds of high $T_{c}$
cuprates.\cite{tsuda,yasuoka,maclaughlin,lombardi}  In the case of undoped
La$_2$CuO$_4$ ($T_N=325$\,K), Tsuda {\it et\,al.}\cite{tsuda} find three
peaks corresponding to the transitions
$|I_{z}$ = $+\frac{3}{2}\rangle
\leftrightarrow
|+\frac{1}{2}\rangle$,  
$|+\frac{1}{2}\rangle
\leftrightarrow
|-\frac{1}{2}\rangle$,
and $|-\frac{1}{2}\rangle
\leftrightarrow
|-\frac{3}{2}\rangle$ for each of the $^{63}$Cu and $^{65}$Cu isotopes
(both have nuclear spin $I=\frac{3}{2}$).  Defining
$A_{x}\equiv A^{ii}$ as the x-component of the on-site hyperfine coupling
constant and $B\equiv A^{ij}$ as the isotropic    supertransferred hyperfine
coupling constant from the four nearest neighbor Cu sites,\cite{mila-rice}
the hyperfine magnetic field  at the nuclear site $i$ can be written as
\begin{equation}
\langle H_{i}\rangle=A_{x}\langle
S_{i}\rangle+B\sum_{j\in nn}^{}\langle 
         S_{j}\rangle
\label{hyperfine1}
\end{equation}
where the sum is taken over the four nearest neighbors.  We also
use the fact that the ordered Cu moments lie within the
ab-plane.\cite{vaknin}  For the  present case of a  two dimensional
square-lattice antiferromagnet, $\langle S_{i}\rangle=-\langle
S_{j}\rangle$ for nearest neighbors $i$ and $j$.   Then Eq.\
\ref{hyperfine1} simplifies to   
\begin{equation}
\langle H_{i}\rangle=(A_{x}-4B)\langle S_{i}\rangle.
\end{equation}
From the analysis of the three transitions of $^{63}$Cu zero field NMR 
(see Fig.~\ref{fig_transition_energies}), Tsuda {\it et\,al.}\ find 
 the internal field at the Cu nucleus to be $\langle
H_{i}\rangle=8.2$\,T in La$_{2}$CuO$_{4}$.
\cite{tsuda} On the other hand, the hyperfine 
coupling constants are known to be $A_{x}=38$ kOe/$\mu_{B}$ and $B=42$
kOe/$\mu_{B}$.\cite{imai93,mila-rice} Together, these values imply an
effective moment of $|\langle S_{i}\rangle|=|\langle
S_{j}\rangle|\approx 0.63\,\mu_{B}$,  in accordance with previous results
of  muon spin precession measurements, which indicate a moment of
$\sim$\,0.58 $\mu_{B}$.\cite{uemura}  Additionally, the frequency
splitting  between two satellite transitions, $|I_{z}$ =
$+\frac{3}{2}\rangle
\leftrightarrow
|+\frac{1}{2}\rangle$
and $|-\frac{1}{2}\rangle
\leftrightarrow
|-\frac{3}{2}\rangle$, provides information regarding the nuclear 
quadrupole interaction tensor, which reflects the second derivative of 
the electrostatic Coulomb potential at the Cu sites.
\cite{abragam,slichter}  The splitting data may be used, 
in principle, to obtain information regarding the charge state of the 
observed nuclei.\cite{tsuda,yasuoka} 

If well defined static charge and magnetic order exist in the striped
materials, we would expect similar, highly structured resonance lineshapes
arising from each unique Cu site, from which we could gain insight into
the spin  state from  the measured hyperfine magnetic field  $H_i$ and the
charge state from the nuclear quadrupole interaction tensor $\nu_Q$. 
Contrary to these expectations,  we observed no evidence for well-defined
periodic spin and charge density modulations even at 350\,mK, as will be
described in the following sections. 

\subsection{$\bf^{63}$Cu NQR/NMR spectra in
La$\bf_{1.875}$Ba$\bf_{0.125}$CuO$\bf_{4}$} 
\label{LBCO_lineshape}

In 1991, magnetic order was discovered in a $x=\frac{1}{8}$ doped sample
of La$_{2-x}$Ba$_{x}$CuO$_{4}$ using $\mu$SR by Luke {\it
et\,al.}\cite{luke} The next year, Tou {\it et\,al.}\
reported anomalous broadening of the low temperature 
$^{63,65}$Cu NQR spectra, consistent with Zeeman perturbed NQR.\cite{tou}
Here we provide a comprehensive  understanding
of the low temperature Cu resonance lineshape in light of the progress 
that has been made in the field of stripe physics.   As demonstrated
earlier by Hunt {\it et\,al.}\cite{hunt} and Singer {\it
et\,al.},\cite{singer}  most of the NQR response of Ba, Nd, and Eu doped
materials is  semi-quantitatively identical for $x\approx\frac{1}{8}$. 
In the following, we focus our discussion of the NQR lineshapes on the 
La$_{1.875}$Ba$_{0.125}$CuO$_{4}$ sample.  The Nd doped materials are ill
suited for Zeeman perturbed NQR measurements at low temperatures  due to
the large Nd$^{3+}$ local moments that make the relaxation  rates too
fast, and the Eu doped samples show extra NQR peaks as discussed in
section \ref{subsec_Eu_lineshape}

\begin{figure}[ht]
\begin{center}
\epsfig{figure=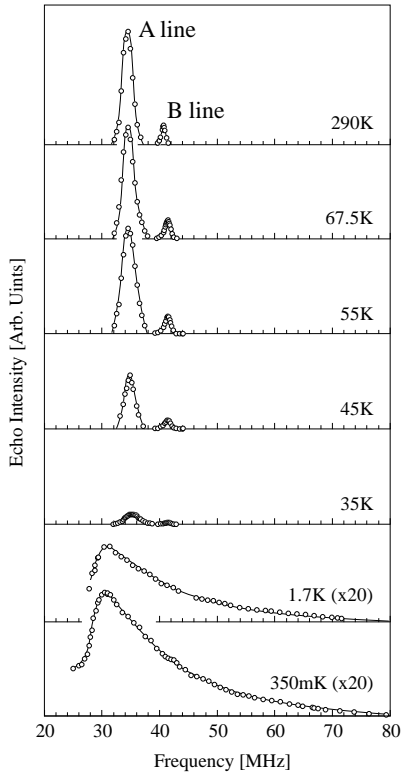,width=2.3in,angle=0}
\end{center}
\caption{$^{63}$Cu NQR lineshapes of a $^{63}$Cu isotope
enriched La$_{1.875}$Ba$_{0.125}$CuO$_{4}$ sample at selected
temperatures, including pure NQR from 290K down to 70\,K, partially
``wiped out" NQR at 55\,K, 45\,K and 35\,K and Zeeman perturbed NQR  at
1.7\,K and 350\,mK.  $^{63}$Cu resonance continues below
$\sim$\,25\,MHz in the  Zeeman pertubed case, but is obscured by the
$^{139}$La NQR
$|I_{z}=\pm\frac{7}{2}\rangle \leftrightarrow |\pm\frac{5}{2}\rangle$
transition peaked near 18\,MHz which is more than an order of magnitude
stronger in intensity (not shown).   Note the scale multipliers when
comparing magnitudes.  The solid curves are guides for the eye.  The
1.7\,K and 350\,mK curves are identical except for overall magnitude.}
\label{fig_LBCO_lineshape}
\end{figure}

Selected $^{63}$Cu NQR lineshapes 
for $^{63}$Cu isotope enriched La$_{1.875}$Ba$_{0.125}$CuO$_{4}$ are shown
in Fig.~\ref{fig_LBCO_lineshape}.  At high temperatures we observe the well
known A and B NQR lines.\cite{yoshimura-imai}  The relative intensity of
the B line is the same as the doped concentration $x$ of the impurity
ions Ba$^{2+}$ and Sr$^{2+}$, over a broad concentration range
$0.02\le x\le0.30$.\cite{yoshimura,imai-unpublished}  Furthermore, the B
line is known to be observable even at elevated temperatures as high as
800\,K.\cite{imai-unpublished}  Accordingly, it is unlikely
that the B line originates from electronic effects such as phase
separation or stripes.  The most likely scenario is that the B line
arises  from Cu sites directly across the apical O from the doped Ba
position and the lattice contribution to the electric field gradient is
different due to the presence of a Ba$^{2+}$ ion rather than
La$^{3+}$.  Needless to say, this line assignment of the B line does not
exclude the possibility that the B site has a somewhat different local
electronic state as compared to the surrounding A sites.  In fact
$^{63}1/T_1$ is known to be somewhat smaller for the B site as shown by
Yoshimura {\it et.\,al.}.\cite{yoshimura-imai}
In Fig.~\ref{fig_Eu_lineshape} we also show that Eu
codoping results in a new structure at the low frequency side of the
A line.  We attribute this new structure to the Cu site sitting across
the apical O from the Eu$^{3+}$ ions.

\begin{figure}[ht]
\begin{center}
\epsfig{figure=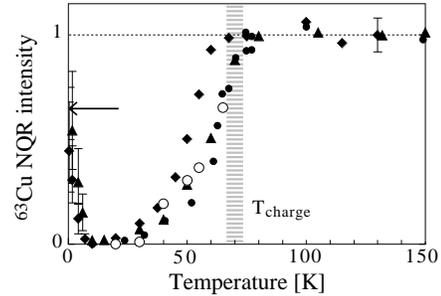,width=2.4in,angle=0}
\end{center}
\caption{Integrated $^{63}$Cu NQR intensity above 25\,MHz for
La$_{1.875}$Ba$_{0.125}$CuO$_4$ (\ding{117}),
La$_{1.68}$Eu$_{0.20}$Sr$_{0.12}$CuO$_4$ ($\blacktriangle$) and 
La$_{1.48}$Nd$_{0.40}$Sr$_{0.12}$CuO$_4$ ($\bullet$)
scaled for the Boltzman factor and $T_2$ spin-echo decay.
In addition, we plot the charge order parameter ($\circ$) deduced from
neutron scattering on
the Nd codoped material.\cite{tranquada1} Within the 
broadened hyperfine field model in Fig.~\ref{fig_simulations_1}f, the
maximum recovery of signal above 25\,MHz is expected to be 61\%, as
indicated by the arrow.}
\label{fig_NQR_intensity}
\end{figure}

\begin{figure}[ht]
\begin{center}
\epsfig{figure=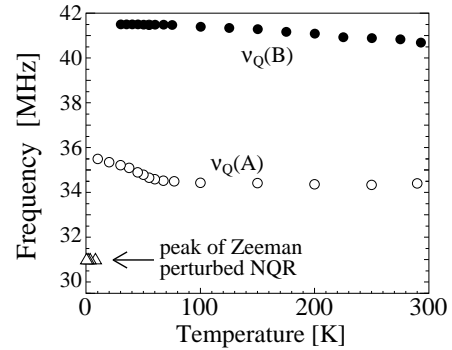,width=2.3in,angle=0}
\end{center}
\caption{The quadrupole resonance frequency, $^{63}\nu_Q$, observed at
fixed pulse separation time 
$\tau=12\,\mu$sec for the A line ($\circ$) and B line ($\bullet$) of a
$^{63}$Cu isotope enriched sample of La$_{1.875}$Ba$_{0.125}$CuO$_{4}$,
as well as the peak frequency of the low temperauture, Zeeman perturbed
NQR spectra ($\triangle$).}
\label{fig_nuQ_plot}
\end{figure}

The temperature dependence of the integrated $^{63}$Cu NQR intensity 
is presented in Fig.~\ref{fig_NQR_intensity}.  Since the occupation 
probability of nuclear spin levels is determined by the Boltzman factor,
the bare NQR intensity is  inversely proportional to temperature $T$. 
Accordingly, we  multiply the bare NQR intensity by $T$ to 
correct for the Boltzman factor.  We also calibrate the intensity 
of the observable NQR signal by extrapolating the observed spin echo
decay to $2\tau=0$. 
The  NQR intensity thus deduced corresponds to the
number of nuclear spins  whose resonance conditions are well defined and
whose relaxation times $^{63}T_{1}$ and $^{63}T_{2}$ 
are long enough to be observed experimentally within the time domain 
of pulsed NQR measurements, $\tau_{NQR}\sim20\,\mu$sec.  Missing signal
intensity arises because at least one of these conditions is violated
for the lost signal.\cite{hunt,singer}      
We note that these are standard
procedures that have also been applied to the NMR investigation of the
SDW state in electron doped Nd$_{2-x}$Ce$_{x}$CuO$_4$ by Kambe {\it
et.\,al.}.\cite{kambe}  Between room temperature and
70\,K the lines retain  constant intensity, as expected, because the
number of $^{63}$Cu nuclear spins  detected by NQR measurements is
unchanged.  The peak frequency, $\nu_Q(A)$, displays only a small
temperature dependence down to $T_{charge}\approx70$\,K as shown in Fig.
\ref{fig_nuQ_plot}.  The weak temperature dependence of $\nu_Q(A)$ down to
70\,K can be  accounted for entirely based on slight tilting of
CuO$_{6}$  octahedra.\cite{singer}  However, as the temperature is
decreased below  70\,K, the temperature dependence of both $\nu_Q(A)$
and $\nu_Q(B)$ changes, and both A and B lines display an identical  drop
in intensity and both nearly disappear by 15\,K.\cite{singer}   This drop
in signal intensity is  ``wipeout''.  As previously noted by
Singer {\it et\,al.}, the estimation of the NQR intensity close to the onset
of wipeout is made more difficult due to the change in the form of the
spin-echo decay.  Given the uncertainties in fitting the spin echo decay,
it is impossible to define
$T_{NQR}$ with accuracy exceeding $\pm$10\%.\cite{singer}  Within these
uncertainties, the onset of wipeout, $T_{NQR}$, is indistinguishable from
$T_{charge}$ for $x\approx\frac{1}{8}$. The wipeout and  subsequent recovery of the intensity below
15\,K will be analyzed  in detail in section \ref{sec_wipeout}.
As the temperature is decreased below   8\,K, the resonant
signal begins to reemerge.  This new line is shown in  the bottom panel of
Fig.~\ref{fig_LBCO_lineshape}.  We note that the character of the line  is
entirely altered.  The peak has shifted down in frequency from 35.5\,MHz
to 31\,MHz. Additionally, the full width at half maximum, which slowly
increases with decreasing temperature  from 2.0\,MHz at 300K to 2.6\,MHz
at 70\,K to 3.2\,MHz at 15\,K, shows sudden and dramatic broadening to
more than 12\,MHz at 350\,mK with the emergence of a large tail towards
higher frequencies.  Note that the spectra at 1.7\,K and 350\,mK are of
identical shape and differ only in magnitude.  In fact, the curves through
those data sets in Fig.~\ref{fig_LBCO_lineshape} are identical except for
vertical scaling.  Based on the analysis presented in the following two
sections, we demonstrate that these changes are consistent with a broad
distribution in the hyperfine magnetic field 
$\langle H_{i}\rangle$ averaged over NQR time scales.

\subsection{Numerical simulations in standard stripe models}
\label{subsec_standard_models}

\begin{figure}[th]
\begin{center}
\epsfig{figure=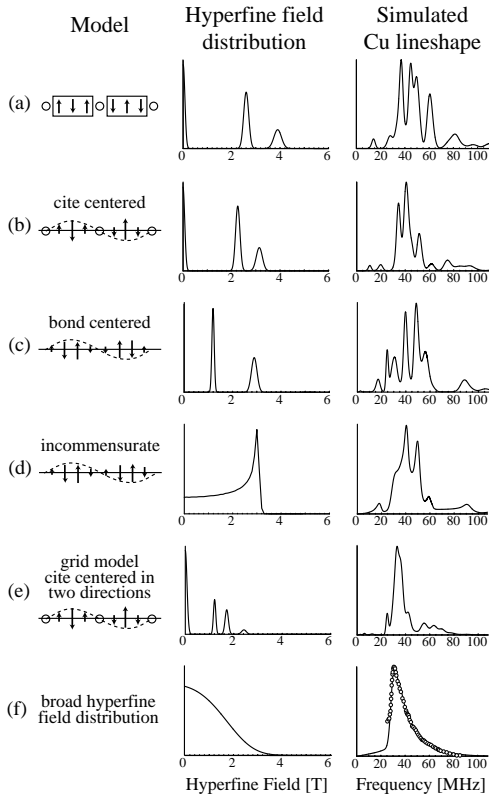,width=2.75in,angle=0}
\end{center}
\caption{Numerical simulations of the $^{63}$Cu spectra based on several
microscopic stripe models.  The first column gives a schematic diagram of
the orientation and magnitude of the Cu magnetic moments along the Cu-O bond
direction.  Arrows and open circles represent ordered Cu moments and a
hole-rich anti-phase boundary without ordered moment, respectively.
 The middle column displays the expected distribution of the
hyperfine magnetic field at the Cu sites.  The right column contains the
corresponding simulated lineshape.  (a)-(c) represent several
commensurate stripe models, and an incommensurate spin density wave is
presented in (d).  Model (e) is a two-dimensional sinusoidal grid
structure  in the CuO$_{2}$ plane and (f) depicts a simple, broad
distribution of hyperfine field that provides a good fit to the
350\,mK La$_{1.875}$Ba$_{0.125}$CuO$_4$ data, which is included as open
circles.}
\label{fig_simulations_1}
\end{figure}

At the time of the discovery  of stripes in Nd doped
La$_{1.48}$Nd$_{0.40}$Sr$_{0.12}$CuO$_{4}$  by neutron
scattering,\cite{tranquada1} Tranquada {\it et\,al.}\ proposed a  simple
model valid at the short timescale of neutron scattering (1\,meV $\approx
2.4\times10^{11}$ Hz, corresponding to 4\,psec) which consists of sets of
three magnetic rows  (a ``stripe'') separated by a charge rich river 
which serves as an antiphase boundary for the antiferromagnetic order. 
This scenario is shown in Fig.~\ref{fig_simulations_1}a.  If the
descriptions of the spin and charge density modulation based on this
model and similar models as shown in Fig.~\ref{fig_simulations_1}b-e,
(which are consistent with the neutron data) are valid
for the entire sample at the much slower NQR timescale of 
$\sim$\,20\,$\mu$sec, we would expect inequivalent Cu sites.  In the case
of  the model of Fig.~\ref{fig_simulations_1}a, we expect there to be
three inequivalent Cu positions: one site in the middle of the magnetic
region, one at the edge and one in the charge rich river. As a first
approximation, we expect  the two magnetic sites at the center and edge of
the stripe to yield three NMR lines each, while the nonmagnetic  boundary
site should produce a single NQR line.  Thus, we might expect a total of
at least seven resonance lines.  These lines are not identifiable in the
data.

Given that the 350\,mK lineshape shown in Fig.~\ref{fig_LBCO_lineshape} is
lacking the expected structure,  we turn to numerical simulations
by solving the Hamiltonian exactly.  Using $I$ as the nuclear spin, $\eta$
as the asymmetry parameter of the electric field gradient (EFG),
$\gamma_n$ as the gyromagnetic ratio, $h$ as Planck's constant, and ${\bf
H}$ as the static hyperfine magnetic field, the Hamiltonian can be written
as\cite{abragam,slichter}

\begin{equation}
{\cal H} = \frac{\nu_{Q}h}{2} \left\{ I_z^2-\frac{1}{3}I(I+1)+\frac{1}{6}
\eta (I_+^2 +I_-^2 ) \right\} - \gamma_n h {\bf H \cdot I},
\label{eq_Hamiltonian1}
\end{equation} 
where the $^{63}$Cu nuclear spin is $I=\frac{3}{2}$ and the
quadrupole resonance frequency $^{63}\nu_Q$, is related to the
electron charge $e$, the maximum of the EFG
$q$, and the
$^{63}$Cu quadrupole moment
$^{63}Q$, by

\begin{equation}
^{63}\nu_Q=\frac{3e^2q}{2hI(I+1)}~^{63}Q
\label{nuQ_equation}
\end{equation}
Because of local axial symmetry of the Cu sites, the electric field gradient
is nearly axially symmetric in both undoped and hole doped CuO$_2$ 
planes.  Additionally, as originally shown by Ohsugi {\it
et\,al.}\cite{ohsugi1,ohsugi2} the $^{139}$La NQR linebroadening takes
place only for
the $|\pm\frac{1}{2}\rangle\leftrightarrow|\pm\frac{3}{2}\rangle$
transition in the low temperature striped phase (also see section
\ref{subsec_La_NQR}).  This indicates that the internal field lies
primarily in the CuO$_2$ plane
for $x\approx\frac{1}{8}$.\cite{nishihara,kitaoka}  Under these
conditions, the Hamiltonian from Eq.~\ref{eq_Hamiltonian1} is greatly
simplified and can be written in matrix form, where we chose the basis
to be the eigentates of the  operator $I_{z}$,
\begin{equation}
{\cal H} = h\left[ 
\begin{array}{cccc}
\frac{-6\gamma_{n}H-\nu_Q}{4} & 0 &
\frac{\sqrt{3}\,\nu_{Q}}{4} & 0 \\ 0 & \frac{-2\gamma_{n}H+\nu_Q}{4} & 0 &
\frac{\sqrt{3}\,\nu_{Q}}{4} \\
\frac{\sqrt{3}\,\nu_{Q}}{4} & 0 & \frac{2\gamma_{n}H+\nu_Q}{4}
& 0 \\ 0 & \frac{\sqrt{3}\,\nu_{Q}}{4} & 0 &
\frac{6\gamma_{n}H-\nu_Q}{4}\\
\end{array}
\right]
\label{Hamiltonian2}
\end{equation}
Between the four  eigenstates of the Hamiltonian, there are six possible
transitions corresponding to six resonant frequencies as shown in
Fig.~\ref{fig_transition_energies}.  Some of these transitions are
nearly  forbidden by the $|\Delta m|=1$ selection rule and exist only due
to the  unusual mixing of  states that occurs in the low and intermediate
field regions where  $\gamma_{n}H \lesssim \nu_{Q}$.  Thus we must
consider the probability for  each transition to occur.  Following G.H.
Muha,\cite{muha} we treat the small experimentally induced R.F. 
field, ${\bf H_{1}}$ ($\sim$\,100\,G), as a perturbation and use the the
standard  time-dependent expression for the transition rate between two
states $n$ and $n'$,

\begin{equation}
P_{n,n'}=|\langle n'| {\bf I\cdot H_{1}} | n \rangle|^{2}.
\label{transition_probability}
\end{equation}
Since we are working on polycrystalline samples where the direction of
${\bf H_{1}}$ with respect to the crystal axis is random, we perform a 
powder average to obtain  the probability for each transition.  These
probabilities correspond to the expected intensity  for each resonance
line in  Fig.~\ref{fig_transition_energies}. Full lineshape simulations 
are computed numerically by summing contributions from  distributions in
both the hyperfine field and in the quadrupole frequency. If the charge
density modulation is sizable,
$^{63}\nu_Q$ may vary depending on the site, and the axial symmetry of
the EFG tensor can break down.  However, the structureless broad
resonance line observed below 8\,K suggests that the spatial modulation
of spin and charge is smooth at the NQR time scale,
$\tau_{NQR}\approx10^{-5}$\,sec.  Hence we believe that approximating the
quadrupole tensor as axially symmetric is acceptable in our simulations. 
Extrapolating  from  observations above 70\,K before wipeout begins we find
$^{63}\nu_Q(A)=34.5$\,MHz for the A site and
$^{63}\nu_Q(B)=41.8$\,MHz for the B site.  Note that, although the
apparent peak frequency of the  A line increases somewhat below 70\,K as
shown in Fig.~\ref{fig_nuQ_plot}, $^{63}\nu_{Q}(A)=35.5$\,MHz at 10\,K is
achieved only for several percent of the sample.  It is not initially
clear what distribution of quadrupole frequencies to employ at low
temperatures, where we expect charge order to induce added width.  In this
section, we take both the A and B lines to be Gaussians, each with a width
of 5.6\,MHz (full width at half maximum), corresponding to twice the value
extrapolated from the A line data above 70\,K.  We discuss the effects of
differing amounts of quadrupole broadening in section
\ref{subsec_broadened_standard_models}.  To allow direct comparisons to 
the data, the simulated intensities are weighted by the square of the
resonant frequency  to  account for experimental sensitivity.\cite{abragam}

The next step is to estimate the distribution of the hyperfine 
magnetic field in various stripe models.  First we examine the
aforementioned model that was initially proposed by  Tranquada {\it
et\,al.},\cite{tranquada1}  which is shown schematically at the left of 
Fig.~\ref{fig_simulations_1}a.  The Cu site in the middle of the  magnetic
``stripe'' is locally in the same enviornment as in the antiferrmagnetic
N\'eel state, so we expect the field at the nucleus to  be
$H^{center}=|A_{x}-4B|\mu_{eff}$ with an an effective magnetic moment of
$\mu_{eff}=0.3\,\mu_{B}$.\cite{luke,nachumi,kojima} Inserting this value
and the hyperfine coupling constants discussed above in section 
\ref{subsec_zeeman_principles}, we estimate $H^{center}=130$\,kOe$/\mu_{B}\times
0.3\,\mu_{B}=3.9$\,T.  For  the Cu site at the edge of the stripe, there
are only three magnetic  nearest neighbor sites that give rise to
supertransferred hyperfine  fields, so the 
expected field is  $H^{edge}=|A_{x}-3B| \mu_{eff}=2.6$\,T.  The 
last Cu site is within the antiphase boundary, where we expect 
oppositely oriented magnetic contributions from two nearest neighbors
which cancel, yielding $H^{boundary}=0$.  We  plot these values of
$H^{center}$, $H^{edge}$ and $H^{boundary}$ in  the distribution of
internal field at the top center of Fig.~\ref{fig_simulations_1}, where
the vertical scale is determined by the  relative numbers of each Cu
site, which is the integral of each peak.  The widths are chosen as
depicted in the figure, where for $H>0$ the widths are proportional to
the hyperfine field $H$ as would be expected for magnetic broadening.  
The numerically computed $^{63}$Cu lineshape is shown at the right of
Fig.~\ref{fig_simulations_1}a.  Agreement between the data and the
simulated lineshape is poor.  Specifically, all of the structures are too
sharp, and even the largest peak is located at $\sim$\,37\,MHz, above the
31\,MHz peak seen in the data.

In addition to the original model proposed by Tranquada {\it et\,al.}, 
we examine site centered and bond centered  commensurate sinusoidal spin
density waves and an  incommensurate sinusoidal spin density wave, as
well as a grid model (site centered sinusoidal spin density wave in two
directions) as shown in Fig.~\ref{fig_simulations_1}b-e.  In each case
$\mu_{eff} = 0.3\,\mu_B$ represents the magnitude of the maximum moment. 
For each model, we follow the same procedure as outlined above  to
construct a hyperfine distribution and compute an expected $^{63}$Cu 
lineshape.  In each of the simulated lineshapes there is too much
structure for a good match to the smooth data.

Even though agreement between the experimental spectra and the 
numerically simulated lineshapes is poor, we notice that, in
Fig.~\ref{fig_simulations_1}d and \ref{fig_simulations_1}e, there is
substantial simulated signal at the experimental peak frequency of
31\,MHz.  These contributions do not arise from the details of the
hyperfine field  distribution, but rather from the nature of the
transition energy  levels themselves.  If we look back to
Fig.~\ref{fig_transition_energies},  we note the two transitions that
start at $\nu_{Q}=34.5$\,MHz and move to lower frequencies with
increasing field.  These lines achieve a broad minimum at 31\,MHz. 
Because of the high density of states that corresponds to resonant
frequencies near 31\,MHz,  {\it any continuous distribution} of internal
fields that includes contributions in the broad window from 0.5\,T to
2.5\,T  will give rise to a lineshape with a 31\,MHz peak.  With this in
mind we examine broad distributions of hyperfine field, and by trial and
error we achieve a good fit to experimental data as shown in
Fig.~\ref{fig_simulations_1}f.  In section
\ref{subsec_broadened_standard_models} we will return to the issue of the
low temperature lineshape to understand the magnetic state of the Cu-O
plane at the slow timescales of our pulsed NQR experiments.

\subsection{Recovered $\bf ^{63}$Cu NQR intensity at 350\,mK}

So far we have considered only the lineshape.  Another important point 
that needs to be addressed is the integrated NQR intensity at 350\,mK.  
The accurate measurement of the recovered integrated NQR intensity below 
8\,K is not easy.  The $^{63}$Cu NQR lineshape measurements above 10\,K can 
be performed in one single scan over frequency without changing the
experimental set up,  but the Zeeman perturbed NQR lineshape is so broad
that we need  to replace the RF coil of the tuned LC-circuit in our NMR
probe three or four times  to cover the entire frequency range from 20 to
90\,MHz.  This enhances  uncertainties in our estimate of the integrated
intensity, and this is the primary reason why we have large error bars for
the integrated  intensity below 8\,K in Fig.~\ref{fig_NQR_intensity}.  In
addition, the lineshape below 25\,MHz is superposed by  an order of
magnitude stronger $^{139}$La  NQR line centered at 18\,MHz.  This does
not allow us to measure  the $^{63}$Cu Zeeman perturbed NQR intensity
below 25\,MHz. Above 70\,K, earlier high field measurements for an aligned
powder sample guarantee that there is no NQR line below
25\,MHz.\cite{imai89} Within the large uncertainties, the recovered signal
intensity  integrated above 25\,MHz represents only 50$^{+30}_{-20}$\% of
the value above 70\,K.  However, it is important to notice that for 
hyperfine fields ranging up to
$\sim$\,4\,T, there are  two branches of Zeeman perturbed NQR transitions
below 25\,MHz in  Fig.~\ref{fig_transition_energies}.  Within our model
calculations of Fig.~\ref{fig_simulations_1}f, that correctly reproduce
the lineshape, we can estimate the fraction of integrated NQR intensity
expected to appear below 25MHz, which is 39\%.  It should be emphasized
that  the echo intensity for the  simulated lineshape plotted in
Fig.~\ref{fig_simulations_1}f is scaled by multiplying by the square of
the frequency to take into account the experimental sensitivity. 
Accordingly, the apparent integrated intensity  below 25\,MHz incorrectly
appears to be a tiny fraction of the whole lineshape.  In  other words,
the maximum integrated intensity expected to reemmerge above 25\,MHz  is
61\% of that at 70\,K, as shown by a horizontal arrow in
Fig.~\ref{fig_NQR_intensity}.  This means that the experimentally
observed intensity  at 350\,mK is effectively 82$^{+49}_{-33}$\% of that
at 70\,K, accounting for nearly the full  intensity from the sample.  This
rules out the possibility that  the descrepancy between the experimentally
observed Zeeman perturbed NQR  lineshape and theoretical calcualtions
based on stripe models originates from missing intensity.  

\subsection{Numerical simulations based on stripe models
including broadening and motionally narrowing}
\label{subsec_broadened_standard_models}

The good agreement between the data at 350\,mK and the simulated lineshape 
and its integrated intensity based on  the broad distribution of
Fig.~\ref{fig_simulations_1}f, indicates that the hyperfine field
averaged over  NMR time scales and over the entire sample should extend
up to about 3\,T  with substantial weight in the region below 2\,T. 
Motivated by the success of the fit for the broad distribution of
hyperfine fields, we return our attention to the stripe models of
Fig.~\ref{fig_simulations_1}a-e to examine the possibility that the
internal hyperfine field corresponding to the stripe models may be
broadened, either by disorder within the sample or by motional averaging
effects, in such a manner as to be consistent with the
$^{63}$Cu NQR spectra.

\begin{figure}[h]
\begin{center}
\epsfig{figure=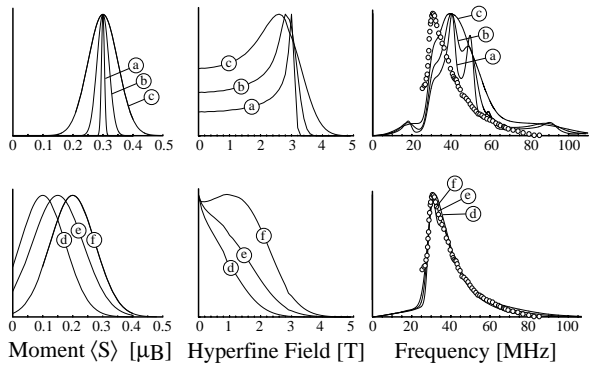,width=3.2in,angle=0}
\end{center}
\caption{Models of incommensurate spin density waves based on a
distribution of maximal moment in different segments of the sample as
given in the leftmost column.  Corresponding to these distributions we
plot the hyperfine field at the nuclear site in the middle column and the
simulated $^{63}$Cu lineshape at the right.  In models (a-c) and (e) we
assume quadrupole width of 5.6\,MHz; in (d) that width is taken to
be 7.5\,MHz and in (f) 3.7\,MHz.  The 350\,mK spectra data is shown as
open circles.}
\label{fig_incommensurate}
\end{figure}

In Fig.~\ref{fig_incommensurate} we develop the hyperfine
field distribution expected from an incommensurate spin density wave,
favored by a recent $\mu$SR study,\cite{kojima} including varying amounts
of broadening, and calculate the corresponding simulated $^{63}$Cu
lineshape.  We take the magnitude of the maximum spin in different regions
of the sample to be
$\langle S\rangle$, which may be distributed due to static
inhomogenieties. We numerically simulate the hyperfine field based on a
row of 10,000 electron spins of magnitude $\langle S_n\rangle=\langle
S\rangle\times\sin(2\pi n/(8+\delta))$ with $|\delta|\ll 1,\delta\neq0$,
taken to be aligned along the $x$ axis and labeled by position $n$.   The
spatial periodicity is that expected from neutron  data.\cite{tranquada1}  
The internal field at each site is computed with the use of
Eq.~\ref{hyperfine1}, giving $\langle H_n\rangle=A_x
\langle S_n\rangle-2B\langle -S_n\rangle-B\langle S_{n+1}\rangle-B\langle
S_{n-1}\rangle$ where we assume N\'eel order in the $y$
direction and use the values of $A_x$ and $B$ as given above.
Distributions arising from different segments of the sample (different
$\langle S\rangle$ values) are combined as per the distribution in
$\langle S\rangle$.  In Fig.~\ref{fig_incommensurate}a-c we show three
simulations based on distributions in the magnetic moment that are
centered at 0.3\,$\mu_B$.  With increasing width in $\langle S\rangle$,
both the hyperfine distributions and the simulated lineshapes become more
smooth.  However, the smoothed lineshaped are shifted to too high a
frequency to match the experimental data.  Thus we rule out the
possibility of a {\it static} incommensurate spin density wave {\it at
NQR time scales}  with maximum magnetic moment of
0.3\,$\mu_B$, even with possible broadening of the value of the moment.

Next we consider the possibilities of a smaller average ordered moment
and of a different amount of quadrupolar broadening.  The simulations of
Fig.~\ref{fig_incommensurate}d-f represent good fits to the data based on
average ordered moments of 0.10, 0.15 and 0.20\,$\mu_B$ paired with
quadrupole widths of 7.5, 5.6 and 3.7\,MHz respectively.
Without this extra quadrupole broadening the contributions of the  A and B
lines give rise to distinct peaks in the simulated lineshape, in
contradiction with our experimental findings. In the
homogeniously doped CuO$_2$ planes of La$_{2-x}$Ba$_x$CuO$_4$ with
$x\gtrsim\frac{1}{8}$ the doping dependence of $\nu_Q(A)$ is
known.\cite{yoshimura} If we imagine the system to consist of many
patches, each of which being homgeniously doped, the extra width in
$\nu_Q(A)$ beyond the extrapolated value from
$T>T_{charge}$ (2.8\,MHz) would correspond to distributions in
$x$ of approximate width $\Delta x\approx0.30$, 0.18 and 0.06
respectively.  There is a trade off between the broadening caused by
magnetic effects and that caused by inherent distribution of the
quadrupole resonance.  However, the simulated lineshape of
Fig.~\ref{fig_incommensurate}e best fits the data, indicating that within
that incommensurate spin density wave model a distribution of
$\langle S\rangle$ centered at 0.15\,$\mu_B$ coupled with moderate
quadrupolar broadening corresponding to $\Delta x\approx0.18$ provides the
best fit to the data, although there is a range of parameters ($\langle
S\rangle$ and quadrupole width) that provide a reasonable fit to the data,
as indicated in Fig.~\ref{fig_incommensurate}d,f.   Of course  this model
assumes patches of homogenious doping rather than microscopic structure
(stripes) which represents an alternative, and perhaps more likely, path
to increased width in $\nu_Q$.

How can we reconcile this result with the conclusion of Kojima {\it
et\,al.}\cite{kojima} based on their $\mu$SR data? Kojima {\it
et\,al.} observed  well defined precession of positive muons 
 consistent with an incommensurate spin density wave with well
defined maximum ordered moment of 0.3\,$\mu_B$.\cite{kojima}  If we
recall that muon spin precession persists for  only about 0.5\,$\mu$sec,
while the NQR spectrum is obtained by spin echo  measurements that detect
hyperfine field averaged over 20\,$\mu$sec, one  possible explanation is
that the relatively large and well-defined moment of 0.3\,$\mu_B$
observed by $\mu$SR measurements\cite{luke,nachumi,kojima} is still
slowly fluctuating during the duration of a Zeeman perturbed NQR
measurent.  Since the hyperfine field is a vector quantity, the averaging
process will always reduce the apparent, averaged field.  Thus, the effect
of motional averaging will be to pull all instantaneous values of the
hyperfine field towards zero,  reducing the effective values.  This may be
the key to resolving the apparant discrepancy between our NQR findings and
the $\mu$SR measurements.  We recall that resistivity never diverges in the
stripe phase,\cite{noda,ichikawa} hence we believe that mobile holes
disrupt spin order.  We note that the motional averaging scenario is
supported by the large value of $^{139}$1/$T_1T$ at 350\,mK (see
Fig.~\ref{fig_1/T1T}), which indicates that low frequency spin
fluctuations persist even at our experimental base temperature.  We also
note that  the motional averaging effects in the present case of Zeeman
perturbed  NQR is somewhat different from the textbook
cases\cite{abragam,slichter} of  {\it motional narrowing} and {\it
exchange narrowing}.   In those cases, one applies a strong and static 
external magnetic field to observe NMR signals.   The applied field
defines the Larmor  precession frequency of nuclear spins, and the dipole
line brodening is only a small  perturbation.  In the present case,
however, the fluctuating  hyperfine field caused by freezing Cu moments is
comparable to or even greater  than the unperturbed NQR frequency.  Thus,
as long as very slow  fluctuations exist in the hyperfine field, the
Larmor precession  itself is not well defined.  Accordingly we expect that
our  Zeeman perturbed NQR spectra is extremely sensitive to very slow 
fluctuations of the hyperfine field. It is worth noting that our case
resembles  the motional narrowing considered by\,Kubo and  Toyabe for zero
or small field resonance\cite{kubo} that has been applied to the analysis
of
$\mu$SR data in glassy systems.\cite{uemura80} 

Although we have focused on the incommensurate spin density wave model
here, the possibilities of static broadening and motional averaging can
be applied to all of the models presented
in Fig.~\ref{fig_simulations_1}.  In each case the sharp structures can be
smoothed out by inhomogeneities within the sample or by motional effects,
and the hyperfine field distributions can be made to be similar to that
of  Fig.~\ref{fig_simulations_1}f, and hence with the experimental data. 
Thus, we cannot exclude any of the  stripe models in
Fig.~\ref{fig_simulations_1} based on our simulations.

\subsection{$\bf^{63}$Cu and $\bf^{65}$Cu resonance spectra in
La$\bf_{1.68}$Eu$\bf_{0.20}$Sr$\bf_{0.12}$CuO$\bf_{4}$ and
La$\bf_{1.64}$Eu$\bf_{0.20}$Sr$\bf_{0.16}$CuO$\bf_{4}$} 
\label{subsec_Eu_lineshape}

\begin{figure}[ht] 
\begin{center}
\epsfig{figure=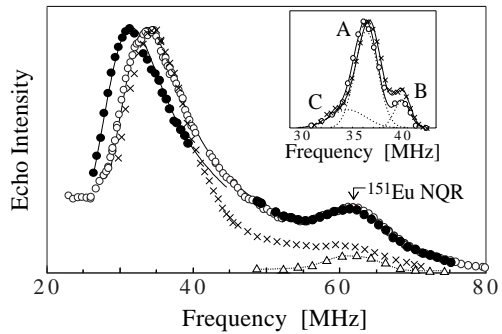,width=2.9in,angle=0}
\end{center}
\caption{Resonance spectra at 1.7\,K 
with a pulse separation time of $\tau=12\,\mu$s for
single isotope enriched samples of
La$_{1.8-x}$Eu$_{0.20}$Sr$_{x}$CuO$_{4}$ ($^{63}$Cu, $x$\,=\,0.12,
$\circ$; $^{65}$Cu, $x$\,=\,0.12, $\bullet$; $^{63}$Cu, $x$\,=\,0.16,
$\times$).  Data taken with longer
$\tau=100\mu$s on the $^{65}$Cu enriched sample is marked by
$\triangle$.  The solid curves represent numerical simulations of the
$x$\,=\,0.12 data as described in the text, while the dotted curve is a
guide for the eye.  Inset: 100\,K lineshapes for the $^{63}$Cu
enriched samples using the same symbols as above.  In the inset the solid
curves represent the sum of three Gaussians representing the A, B
and C peaks, which are individually depicted by dashed curves for the
$x$\,=\,0.12 sample.  For both cases the intensity ratio
A:B:C is consistent with the expected form
0.8\,-\,$x$\,:\,$x$\,:\,0.2 from Cu sites away from all dopants, closest
to the Sr ions and next to the Eu dopants, respectively.}
\label{fig_Eu_lineshape}
\end{figure}

In Fig.~\ref{fig_Eu_lineshape} we show the Zeeman perturbed NQR spectum
for two samples of La$_{1.68}$Eu$_{0.20}$Sr$_{0.12}$CuO$_{4}$, one
enriched with the $^{63}$Cu isotope and the other with $^{65}$Cu.  The
peak frequency  of the major peak is 34$\pm$0.75\,MHz for the $^{63}$Cu
enriched sample and 31$\pm$0.75\,MHz for the sample enriched with
$^{65}$Cu.  The peak  frequency of the $^{63}$Cu sample is
3$\pm$1.5\,MHz higher than the 31\,MHz observed for $^{63}$Cu in
La$_{1.875}$Ba$_{0.125}$CuO$_4$.  The shift of the Zeeman perturbed NQR
peak between these two materials arises primarily from the difference in
the intrinsic quadrupole frequencies, which is 2.0\,MHz at 70\,K. 
Additionally, the greater proximity of the B line in the Eu codoped
material (see the inset of Fig.~\ref{fig_Eu_lineshape}) contributes to a
higher peak frequency in the Zeeman perturbed spectra, whereas the
greater separation of the B line in La$_{1.875}$Ba$_{0.125}$CuO$_4$
results in increased width.

The only extra feature visible in these lineshapes for the Eu doped
materials as compared with those of La$_{1.875}$Ba$_{0.125}$CuO$_{4}$
(recall Fig.~\ref{fig_LBCO_lineshape}), is the peak at 62\,MHz.  We also
observed extremely weak signals that  continue above 80\,MHz.  We believe
that these signals arise from NQR of the Eu nuclei themselves.  We note
that the peak at 62\,MHz is observed only in the Eu doped materials and
is independent of the Cu isotope present in the $^{63,65}$Cu isotope
enriched samples.  The relaxation time at the 62\,MHz peak is also slow,
allowing us to use increased delay time $\tau$ to separate it as shown by
open triangles in Fig.~\ref{fig_Eu_lineshape}a.  The reduced relaxation
rate is consistent with the small hyperfine field expected at the Eu
site, which is substituted at the La position.  Furthermore, if we assume
that the electric field gradient is the same at the Eu and La positions,
use the ratio of the quadrupole moments for the two nuclei
($^{151}Q/^{139}Q$ = 4.5), and employ Eq.~\ref{nuQ_equation}, we find that
the expected peak position of $^{151}$Eu NQR is $^{151}\nu_Q=59.5$\,MHz,
very similar to our observation.  Finally, we note that the $^{151}$Eu
peak is still observed at 77\,K, strongly indicating that there is no
connection with the low temperature Cu spectra.

Disregarding the Eu NQR signal, we focus on the Cu resonance
in La$_{1.68}$Eu$_{0.20}$Sr$_{0.12}$CuO$_{4}$ below 55\,MHz.  
We repeat the same lineshape analysis as discussed above in section
\ref{subsec_standard_models} for the Eu doped materials.  We 
substitute $^{63}\nu_{Q}(A)=36.2$\,MHz,
$^{63}\nu_{Q}(B)=40.2$\,MHz and $^{63}\nu_{Q}(C)=34.5$\,MHz as
extrapolated from temperatures above the onset of wipeout and include the
slightly greater ($\sim$\,25\% greater than the 5.6\,MHz used in
Fig.~\ref{fig_simulations_1}) width in quadrupole frequencies that is
characteristic of the Eu doped materials.  We employ the same distribution
of hyperfine magnetic field as   above for
La$_{1.875}$Ba$_{0.125}$CuO$_{4}$ (see Fig.~\ref{fig_simulations_1}e). 
For the $^{65}$Cu isotope enriched sample we use the ratio of qudrupole
moments ($^{65}Q/^{63}Q=0.924$) and the gyromagnetic ratios
($^{65}\gamma/^{63}\gamma_n=1.071$) between isotopes to determine
$^{65}\nu_{Q}(A,B)$ and to scale the response to the hyperfine field.  In this way,
there are no free parameters in the fit of either the $^{63}$Cu
or the $^{65}$Cu spectra. The numerically calculated lineshapes are shown
in  Fig.~\ref{fig_Eu_lineshape}a as solid lines. The good agreement
between these simulated lineshapes with the experimental data for both
isotopes supports the Cu spectra analysis that has been presented
throughout section \ref{section_zeeman_perturbed}.

In order to investigate the influence of different hole concentrations
on Zeeman perturbed $^{63}$Cu NQR spectra, we carried  out preliminary
measurements in $^{63}$Cu isotope enriched
La$_{1.64}$Eu$_{0.20}$Sr$_{0.16}$CuO$_{4}$.  The recovered integrated
intensity at 1.7\,K is more than a factor of two smaller than  in
La$_{1.68}$Eu$_{0.20}$Sr$_{0.12}$CuO$_{4}$ (See
Fig.~\ref{fig_anchor_points}c), in agreement with  Teitelbaum {\it
et\,al.}\cite{teitelbaum}   The overall lineshape presented in
Fig.~\ref{fig_Eu_lineshape} is very similar to the case of 
La$_{1.68}$Eu$_{0.20}$Sr$_{0.12}$CuO$_{4}$, but the high frequency tail
is smaller and the width of the  main peak 10.5$\pm$0.5\,MHz, is narrower
than 12.5$\pm$0.5\,MHz in La$_{1.68}$Eu$_{0.20}$Sr$_{0.12}$CuO$_{4}$.  
Within the lineshape analysis presented above in section
\ref{section_zeeman_perturbed}, this change in width implies that the 
maximum magnitude of the hyperfine field in 
La$_{1.64}$Eu$_{0.20}$Sr$_{0.16}$CuO$_{4}$ is 5-10\% smaller than in
La$_{1.68}$Eu$_{0.20}$Sr$_{0.12}$CuO$_{4}$ at 1.7\,K.  There  are a few
possible explanations for the difference in hyperfine field.  First,
since stripes are less stable at  hole concentrations
$x\neq\frac{1}{8}$, \cite{singer,teitelbaum,kumagai,ohsugi1} there may be
additional  fluctuations for those concentrations that may motionally
narrow the main peak.  This picture  is consistent with the fact that at
1.7\,K, a smaller fraction of the Cu  signal intensity reemerges as Zeeman
perturbed NQR in the samples away from $x\approx
\frac{1}{8}$.\cite{teitelbaum}  Second, it is possible that the
increased disorder caused by the additional holes may increasingly
frustrate the spins, and reduce the magnitude of ordered 
moments.  We also note that the peak is slightly higher in the $x=0.16$
material as compared to $x=0.12$, because $\nu_Q(A)$ increases with hole
doping.

Before closing this section we would like to comment breifly on
two recent reports of low temperature $^{63}$Cu NQR spectra in
La$_{1.8-x}$Eu$_{0.20}$Sr$_{x}$CuO$_{4}$\cite{teitelbaum} and
La$_{1.875}$Ba$_{0.125}$CuO$_{4}$.\cite{teitelbaum2}  Both of these
publications show broad lineshapes superposed with what
appears to be NQR signal from the high temperature A and B lines that have
survived to low temperature.  We have seen comparable results on a poorly
annealed sample of La$_{1.875}$Ba$_{0.125}$CuO$_{4}$ in which wipeout was
not complete.  After careful reannealing of this sample, wipeout became
complete and the A and B lines disappeared from the low temperature
spectra.  We note that potential problems with the lineshapes does not
diminish the primary finding of Teitel'baum {\it et\,al.}\ regarding the
doping dependence of the recovered signal, i.e.~that the intensity of the
recovered signal at 1.3\,K in La$_{1.8-x}$Eu$_{0.20}$Sr$_{x}$CuO$_{4}$ is
greatest for $x\approx\frac{1}{8}$, indicating the increased stability of
that doping level.\cite{teitelbaum}   This conclusion is in agreement
with our earlier finding that the  temperature dependence of wipeout
effects in La$_{1.6-x}$Nd$_{0.40}$Sr$_{x}$CuO$_{4}$ is sharpest for
$x\approx\frac{1}{8}$.\cite{singer}  It is also  consistent with another
earlier finding based on hyperfine broadening of the $^{139}$La NQR line
at 1.3\,K by Ohsugi {\it et\,al.}\ indicating that stripes are most
stable near  $x\approx0.115$ in La$_{2-x}$Sr$_{x}$CuO$_{4}$.
\cite{ohsugi1}

\subsection{$\bf^{139}$La NQR lineshapes in
La$\bf_{1.875}$Ba$\bf_{0.125}$CuO$\bf_{4}$} 
\label{subsec_La_NQR}

To this point we have discussed only the results of $^{63,65}$Cu NQR, but
resonance studies of the $^{139}$La site can also provide additional
insight.\cite{ohsugi1,ohsugi2,nishihara} Since the $^{139}$La
nucleus has spin $I=\frac{7}{2}$, the $^{139}$La NQR spectra of a single
site consists of three peaks, roughly corresponding to 1, 2 and 3 times
$^{139}\nu_Q$.  With the addition of a  perturbing magnetic field,
the three resonance lines are split in a manner that depends on the
details of the direction and the magnitude of the hyperfine
field.\cite{nishihara} In earlier $^{139}$La NQR studies of the low
temperature striped phase of  La$_{2-x}$Sr$_x$CuO$_4$, Ohsugi {\it
et\,al.}\ found that the $|\pm\frac{1}{2}\rangle\leftrightarrow
|\pm\frac{3}{2}\rangle$ transitions at $\sim$\,6\,MHz broadens for
$x\approx \frac{1}{8}$ at 1.3\,K, while the
$|\pm\frac{5}{2}\rangle\leftrightarrow |\pm\frac{7}{2}\rangle$ 
transitions at $\sim$\,18\,MHz do not show magnetic line
broadening.\cite{ohsugi1,ohsugi2}  This  implies that the  time-averaged
hyperfine field lies  within the CuO$_2$ plane.\cite{nishihara,kitaoka} 
In this case, the extra line brodening is roughly equal to 
$4\gamma_nH_{ab}$, where $H_{ab}$ is the hyperfine field.

\begin{figure}[ht]
\begin{center}
\epsfig{figure=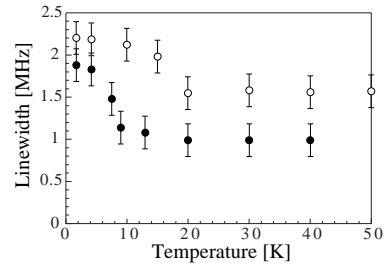,width=2.3in,angle=0}
\end{center}
\caption{$^{139}$La linewidth as a function of temperature 
for the $|\pm\frac{1}{2}\rangle\leftrightarrow
|\pm\frac{3}{2}\rangle$ transition at a frequency $^{139}\nu_{Q} 
\approx$ 6\,MHz for La$_{1.875}$Ba$_{0.125}$CuO$_{4}$ ($\bullet$) and
La$_{1.68}$Eu$_{0.20}$Sr$_{0.12}$CuO$_{4}$ ($\circ$).} 
\label{fig_La_width} 
\end{figure}

\begin{figure}[ht]
\begin{center}
\epsfig{figure=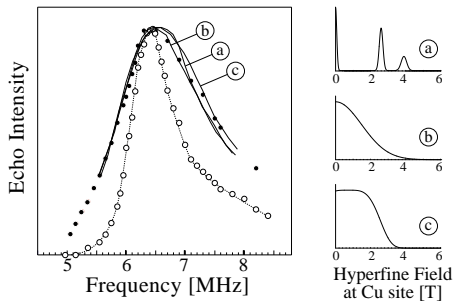,width=2.5in,angle=0}
\end{center}
\caption{$^{139}$La NQR spectra of
La$_{1.875}$Ba$_{0.125}$CuO$_{4}$ for the $|\pm\frac{1}{2}\rangle\leftrightarrow
|\pm\frac{3}{2}\rangle$ transition at 20\,K ($\circ$) and
1.7\,K ($\bullet$) in the main plot to the left.  The solid lines
represent simulated lineshapes that arise from three different
distributions of internal hyperfine magnetic field as discussed in the
text.  The hyperfine distributions are depicted in the small plots to the
right, and are labeled (a-c) to match the corresponding simulated
lineshape.}
\label{fig_La_NQR_LBCO} 
\end{figure}

Our goal is to use the broadening that we observe in the
$|\pm\frac{1}{2}\rangle\leftrightarrow |\pm\frac{3}{2}\rangle$  
$^{139}$La NQR transition as a test of the models presented above in
sections \ref{subsec_standard_models}.  We present the temperature 
dependence of the linewidth of the $|\pm\frac{1}{2}\rangle\leftrightarrow
|\pm\frac{3}{2}\rangle$  $^{139}$La NQR transition in
Fig.~\ref{fig_La_width} for  La$_{1.875}$Ba$_{0.125}$CuO$_{4}$.
Lineshape data from the same sample is presented in
Fig.~\ref{fig_La_NQR_LBCO}.  The linewidth broadens from
1.0\,MHz at 20\,K to 1.9\,MHz at 1.7\,K, but does not display the clear
splitting that is observed in the N\'eel state of La$_{2}$CuO$_{4}$.
\cite{maclaughlin,nishihara,kitaoka}  Similar temperature dependence was
found in La$_{1.68}$Eu$_{0.20}$Sr$_{0.12}$CuO$_4$ as shown in
Fig.~\ref{fig_La_width}.
Increase of linewidth at a somewhat higher temperature implies
that slowing of spin fluctuations is more rapid in
La$_{1.68}$Eu$_{0.20}$Sr$_{0.12}$CuO$_4$ than in
La$_{1.875}$Ba$_{0.125}$CuO$_{4}$, consistent with the larger
$^{139}1/T_1T$ at the same temperature in the former material, as seen in
Fig.~\ref{fig_1/T1T}b. In a recent work, Teitel'baum {\it
et.\,al.}\ also found similar behavior for a Nd doped sample with
$x=0.12$.\cite{teitelbaum-archive}  

To test the consistency of the
$^{139}$La line broadening with the models proposed above to account for
the $^{63}$Cu Zeeman perturbed lineshape, we undertake an analysis very
similar to that of sections
\ref{subsec_standard_models}.  Again we perform an exact diagonalization
of the Hamiltonian (Eq.~\ref{eq_Hamiltonian1}), but this time for spin
$I=\frac{7}{2}$ employing the same  value of $\eta$ and the
same small tilting of spin orientation as deduced by  Nishihara {\it
et\,al.}.\cite{nishihara}  We confirmed that our conclusions do not 
depend on small variations in these additional parameters.
In the right hand column of Fig.~\ref{fig_La_NQR_LBCO}, we present
three model hyperfine field distributions including (a) and (b) which
are the same hyperfine field distributions as used in the $^{63}$Cu
lineshape analysis as shown in Fig.~\ref{fig_simulations_1}a and
\ref{fig_simulations_1}f.  To avoid confusion,
the magnitude of the hyperfine field is still specified by its value at
the $^{63}$Cu site, although its value is greatly decreased at the
$^{139}$La position because of the two orders of magnitude smaller
hyperfine  coupling constant.  Specifically, we take the ratio of the
hyperfine field at the $^{139}$La and $^{63}$Cu sites to be the same as in
undoped La$_2$CuO$_4$, which is 0.1\,T/8.2\,T=0.012.\cite{kitaoka,tsuda}  
This means that if there is 4\,T of hyperfine field at a Cu site,  there
is 0.048\,T of hyperfine field at nearby La sites.  We must  caution that
this ratio can, in principle, be dependent on the exact material chosen
and especially on low temperature structural transitions which alter the
mixing of atomic orbitals.  In addition, if there is a  strong modulation
of the hyperfine field at nearby Cu sites, the La ions may  experience an
average of those values.  However, we find that the use of this  ratio
0.012 adequately describes our data.  Finally, we take the 20\,K
$^{139}$La NQR data of Fig.~\ref{fig_La_NQR_LBCO} and numerically simulate
the added width that arises from the static hyperfine field distributions
given by the distributions (a-c) of Fig.~\ref{fig_La_NQR_LBCO}.  These
simulated low temperature lineshapes are depicted as solid lines in the
main panel of Fig~\ref{fig_La_NQR_LBCO}.  We note that the differences
between the simulated lineshapes corresponding to the different hyperfine
field distributions are small, indicating that it is impossible to choose
one over another based on the $^{139}$La NQR measurement alone.  However,
we also note that all are consistent with the observed broadening of the
$^{139}$La NQR lineshape, giving added weight to the correctness of our
explanation of the $^{63,65}$Cu and $^{139}$La spectra in
La$_{1.875}$Ba$_{0.125}$CuO$_4$ as presented throughout this section.

\subsection{Summary of Zeeman perturbed NQR lineshape measurements and
analysis}

We have demonstrated that for La$_{1.875}$Ba$_{0.125}$CuO$_{4}$,
La$_{1.68}$Eu$_{0.20}$Sr$_{0.12}$CuO$_{4}$ and
La$_{1.64}$Eu$_{0.20}$Sr$_{0.16}$CuO$_{4}$, the $^{63}$Cu NQR signal that is wiped
out at high temepratures reemerges below $\sim$\,8\,K as a single, broad
peak with a lower peak frequency.  The broad linewidth is caused by
hyperfine magnetic  fields from frozen Cu magnetic moments that are
nearly static over NQR  time scales.  The hyperfine field $H$ is
primarily  within the CuO$_{2}$ planes and the  hyperfine fields extend up
to $3-4$\,T.  The  upperbound of the field distribution corresponds to
the maximum frozen Cu magnetic moments of 0.2$\mu_{B}-0.3\mu_{B}$, in
rough agreement with  estimates based on $\mu$SR.  However, the field
distribution is not  consistent with the sharply defined stripe models
that properly account for the  elastic neutron scattering data and the
precession of muons at faster time scales.  

To fit the  Zeeman perturbed NQR within the stripe models, including the
incommensurate spin density wave  picture that is favored by $\mu$SR, the
observed  lineshape requires a field distribution centered close to zero
with very large distribution coupled with an increased quadrupole width. 
The magnitude of corresponding maximum  Cu magnetic moments,
$\sim$\,0.15$\,\mu_{B}$, is suppressed relative to the
$\sim$\,0.3$\,\mu_{B}$ as determined by $\mu$SR studies.  This,  together
with somewhat narrower lineshape observed for
La$_{1.64}$Eu$_{0.20}$Sr$_{0.16}$CuO$_{4}$ and relatively high 
relaxation rate $^{139}1/T_{1}T$ observed even at 350\,mK,
indicates that motional narrowing that is caused by remnant slow 
fluctuations which persist even at our base temperature.  The presence
of slow fluctuations has also been emphasized by Suh
{\it et\,al.}.\cite{suh-rapid}  The fact that the 1.7\,K and 350\,mK
spectra are of identical shape suggests that the decreasing frequency
scale of charge and spin fluctuations has saturated and that there are
many low lying excited states of the stripes, while the continuing NQR
signal recovery may indicate that an increasing fraction of the sample is
experiencing the saturated, low frequency fluctuations.

The increased quadrupole width that is required to smooth the A and B
lines into the low temperature lineshape indicate the presence of a
broadly distributed charge environment which alters the electric field
gradient site by site.  Although we find in section \ref{section_slowing}
that we can phenomenologically account for the $^{63}$Cu wipeout data
based on slowing spin fluctuations, this quadrupole broadening indicates
that spatial variation of the local charge environment does indeed exist
throughout the sample.

\section{Slowing of Fluctuation Timescale for the $\bf\frac{1}{8}$ phases}
\label{section_slowing}
\label{sec_wipeout}

\subsection{Preliminary considerations on the fluctuation frequency scales
of  stripes}

In general, the growth of short range antiferromagnetic spin-spin
correlations causes the spin fluctuation frequency scale $\Gamma=1/\tau_{spin}$ 
to gradually slow down.  In the normal metallic state above $T_{c}$ in
the high $T_{c}$ cuprates, the  characteristic fluctuation frequency
$\Gamma$  of spin fluctuations decreases roughly in proportion to the
temperature $T$.   This peculiar slowing down is frequently referred to
as ``$\omega/T$ scaling,"  because the frequency dependence of
$\chi''({\bf q},\omega)$ satisfies a simple scaling law when the
frequency $\omega$ is normalized by the temperature $T$ (or 
equivalently, by $\Gamma$).\cite{keimer,sternlieb}  One can  define
$\Gamma$ as the peak frequency of the $\omega$ dependence  of
$\chi''({\bf q},\omega)$, or, more precisely, by fitting $\chi''({\bf
q},\omega)$ to  a relaxational function with appropriate frequency
dependence.   We fit $\chi''({\bf q},\omega)$ observed by Aeppli {\it
et\,al.}\cite{aeppli} for La$_{1.86}$Sr$_{0.14}$CuO$_{4}$ to a Lorentz
oscillator form to deduce $h\Gamma\approx8.8$\,meV at 35\,K.  By scaling
the magnitude of $h\Gamma$ obtained at 35\,K for 
La$_{1.86}$Sr$_{0.14}$CuO$_{4}$ linearly with temperature, we estimate
$h\Gamma\approx17.6$\,meV at $T_{charge}=70$\,K.  Needless to say,
$h\Gamma$ should depend on the hole  concentration $x$ (for example,
$h\Gamma\approx4$\,meV at similar temperatures for
$x=0.04$\cite{keimer}).  We expect a somewhat smaller value of $h\Gamma$
for $x=0.12$ than for $x=0.14$  because the dynamic spin-spin
correlation  length is somewhat longer for $x=0.12$ than for
$x=0.15$.\cite{yamada} However, since $^{63}1/T_{1}$ depends  little on
$x$ for $0.12\leq x\leq0.20$, we do not expect a major change in  the
magnitude of $h\Gamma$.  We note that $^{63}1/T_1$ in
La$_{1.8-x}$Eu$_{0.2}$Sr$_x$CuO$_4$ is nearly identical with that of
La$_{2-x}$Sr$_x$CuO$_4$.  Hence, it is safe to assume that the spin
dynamics above $T_{charge}$ depend little on Eu or Nd rare earth codoping.

The important point to notice is that this mild T-linear slowing down of
spin  fluctuations with temperature breaks down in the charge ordered
segments of the CuO$_{2}$ planes below $T_{charge}\approx70$\,K. 
EPR measurements by Kataev and
co-workers\cite{kataev} indicate that spin fluctuations are slowing
more quickly below $T_{charge}$.   Somewhat below the onset of charge
ordering, spin fluctuations in some segments of the
La$_{1.48}$Nd$_{0.4}$Sr$_{0.12}$CuO$_{4}$   sample become slow enough at
$T_{spin}^{neutron}=50$\,K to be considered static, and neutron
scattering  measurements begin to detect quasi-elastic scattering with
excitation energy $\hbar\omega\approx1$\,meV.  We emphasize that this does
not mean that  all the Cu spins have slowed down to
$\hbar\Gamma\approx1$\,meV.  The  fact that some Cu NQR signal is still
observable even below $T_{charge}$, and the
$^{63}1/T_1$ measured for the observable segments  does not show any
anomalous behavior\cite{hunt,singer,imai89,imai93} indicates that the
observable Cu NQR signals originate from other segments of the CuO$_{2}$
plane where $\Gamma$ is still smoothly slowing following the $\omega/T$ 
scaling. One may imagine that phase separation sets in with
microscopic length scales below $T_{charge}$, and $\Gamma$
begins to exhibit glassy slowing with a broad distribution because
$\Gamma$ slows down exponentially in the charge ordered segments.
This is in sharp contrast with the narrow and well defined
spin fluctuation frequency
$\Gamma$ in the  ordinary metallic state of CuO$_{2}$ planes above
$T_{charge}$.  We also note that if charge order is a second order phase
transition, the critical slowing of charge dynamics should exist even
above $T_{charge}$.  In fact, the resistivity data does show an upturn
prior to $T_{charge}$.\cite{noda,ichikawa}  This means that
precursors of glassy slowing may exist at temperatures slightly above
$T_{charge}$.  Our earlier observation that the Gaussian component of
$^{63}1/T_2$ shows gradual crossover to Lorentzian due to motional
narrowing $\sim$\,10-30\,K above $T_{charge}$ supports
this idea.

The inelastic magnetic neutron scattering intensity reaches 
about 80\% of the maximum value by 25\,K, meaning that 80\% 
of the spins are slowed to below the time scale
of $\tau_{neutron}\approx10^{-11}$\,sec by 25\,K.  To our knowledge, no
detailed inelastic neutron scattering data is available for
La$_{2-x-y}$(Nd,Eu)$_{y}$Sr$_{x}$CuO$_{4}$ with
$x\approx\frac{1}{8}$  near $T_{charge}$ and $T_{spin}$ other than a
limited data set reported by Tranquada {\it et\,al.}.\cite{tranquada4} 
Accordingly, the details of how the slowing of spin  fluctuations deviate
from the high temperature ``$\omega/T$ scaling" towards
$\hbar\Gamma\approx1$\,meV at 50\,K as a function of  temperature is
unknown.  In what follows we smoothly extrapolate the low temperature
behavior to the $\omega/T$ curve.

In this temperature range, $\mu$SR measurements do
not detect any  static hyperfine fields.  This means that Cu spins fluctuate so quickly 
that the hyperfine field is averaged to zero.  
With decreasing temperature, $\mu$SR measurements begin to 
detect static hyperfine magnetic field that last longer than 
0.1\,$\mu$sec between 25 and 30\,K.  This means that the spin 
fluctuation frequency $\Gamma$ has slowed down to $\sim$10$^{7}$ Hz 
in some segments of the CuO$_{2}$ planes.  
At 30\,K, the Cu NQR signal is almost completely wiped out as shown in
Fig.~\ref{fig_NQR_intensity}.  This  indicates that normal metallic
segments of CuO$_{2}$ planes no longer  exist below 30 K.  By extending
the $\omega/T$ scaling down to 30 K,  we estimate the upper limit of the
distribution of spin fluctuation  frequency $\Gamma\sim10^{12}$ Hz at 30
K.  On the other hand, the  lower limit from the slowest components of
spins is $\sim$10$^{7}$ Hz.

\begin{figure}[ht]
\begin{center}
\epsfig{figure=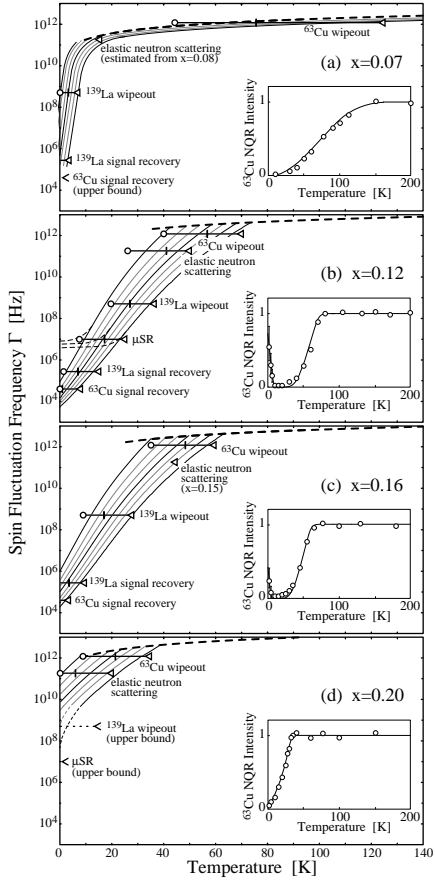,width=2.6in,angle=0}
\end{center}
\caption{The distributions of the spin fluctuation frequency,
$\Gamma$, as a function of temperature for Nd and Eu codoped
La$_{2-x}$Sr$_{x}$CuO$_4$ for several values of $x$.  For each
measurement, the triangles, the small vertical line segments and the open
circles correspond to the onset, the 50\% point and the 80\% point of the
anomaly, respectively.  In some cases data is not available and we
estimate values from similar samples, as noted on the figure.  The
vertical positioning of the points indicates the appropriate timescale
for each measurement, as discussed in section \ref{subsec_wipeout}.  The
thick dashed line across the top of each panel represents $\omega/T$
scaling where there is little distribution of $\Gamma$.  The experimental
data serves as anchoring points for the distribution of $\Gamma$
that is used in the wipeout calculations, which is depicted as a contour
plot utilizing the solid black and gray curves.\cite{note-log-scales}  The
region between any two adjacent curves represents 12.5\% of the volume of
the sample, with an additional 12.5\% above the top curve and below the
lowest.  Dashed lines near
$\Gamma\approx6\times10^6$\,Hz for $x=0.12$ represent the distribution
$\Gamma$ deduced in the Gaussian model as discussed in the text.  The
insets show the
$^{63}$Cu wipeout data and the simulated wipeout based on
Eq.~\ref{eq_wipeout_fraction}.}
\label{fig_anchor_points}
\end{figure}

The line broadening of Zeeman perturbed NQR caused by static hyperfine 
magnetic fields takes place for $^{139}$La NQR and $^{63,65}$Cu at an even
lower  temperature, $T_{NQR}^{Zeeman}\approx$ 18\,K and 8\,K,
respectively (see Figs.~\ref{fig_NQR_intensity} and
\ref{fig_La_width}).   This indicates that the fluctuation frequency
spectrum has slowed to a timescale comparable to the duration of a spin
echo experiment. This time scale is set by  the separation of two RF
pulses, usually $\tau=12\,\mu$sec, i.e. if spins   fluctuate much slower
than $h\Gamma\approx10^{5}$\,Hz, the  hyperfine field would look entirely
static and we expect full  hyperfine broadening.  As explained in section
\ref{section_zeeman_perturbed}, our Zeeman  perturbed NQR results seem to
indicate that this condition is not  quite satisfied in the present case
even at 350\,mK.  In Fig.~\ref{fig_anchor_points} we present the summary
of experimental information regarding the distribution of slowing spin
fluctuations.

In the following sections, we will demonstrate that the glassy 
slowing of spin fluctuations below $T_{charge}$ for $x\approx\frac{1}{8}$ 
can be explained consistently based on  the renormalized classical
scaling of the  Non-Linear-$\sigma$ model with a reduced value of spin
stiffness $2\pi\rho_{s}^{eff}$.  In section \ref{subsec_spin_stiffness},
we estimate the numerical value $2\pi\rho_{s}^{eff}\approx200$ K based on
a simple physical argument.  We will also explain why the {\it indirect}
spin contribution to wipeout can be triggerred by charge ordering.   In
section \ref{subsec_T1}, we analyze the sudden and  dramatic increase of
$^{139}1/T_{1}T$ and its distribution  that takes place below 
$T_{charge}$ based on the renormalized classical scaling, and deduce 
the spatial distribution of $\Gamma$ as a function of temperature.  In 
section \ref{subsec_wipeout}, we analyze $^{63}$Cu and $^{139}$La NQR
wipeout effects  together with magnetic neutron scattering and $\mu$SR
data, again  based on the renormalized classical scaling.  
The distribution of $\Gamma$ deduced by this approach agrees
well with that deduced from $^{139}1/T_{1}$.

\subsection{Spin stiffness in charge ordered segments 
and its relation to $\bf^{63}$Cu NQR wipeout}
\label{subsec_spin_stiffness}

\begin{figure}[ht]
\begin{center}
\epsfig{figure=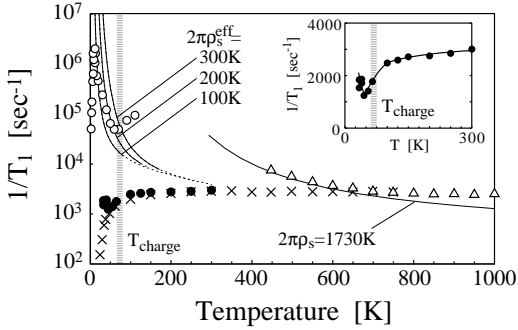,width=3.23in,angle=0}
\end{center}
\caption{$^{63}1/T_1$ data as a function of temperature for undoped
La$_2$CuO$_4$\cite{imai93} ($\triangle$),
La$_{1.85}$Sr$_{0.15}$CuO$_4$\cite{imai93} ($\times$), and
La$_{1.68}$Eu$_{0.20}$Sr$_{0.12}$CuO$_4$\cite{singer} ($\bullet$).  We
also show $^{139}1/T_1$ scaled by Eq.~\ref{eq_T1_ratio} for the Eu codoped
material ($\circ$).  This quantity roughly represents $^{63}1/T_1$ of
the unobservable $^{63}$Cu NQR signal in the stripe ordered segments.  The
curves represent $^{63}1/T_1$ in the renormalized classical scaling
model with different values of $2\pi\rho_s^{eff}$ as listed.  Inset:
the same data for the Eu codoped material on a linear scale for clarity.}
\label{fig_combo_T1}
\end{figure}

The $^{63}$Cu nuclear spin-lattice relaxation rate $^{63}1/T_{1}$ diverges
exponentially in the renormalized  classical scaling regime of both
isotropic\cite{imai93} and  anisotropic\cite{thurber-3leg}  quasi-2d
Heisenberg systems.   As shown in Fig.~\ref{fig_combo_T1},  the
exponential divergence of $^{63}1/T_{1}$ was successfully observed in 
undoped La$_{2}$CuO$_{4}$, and the fit to the renormalized classical
expressions allowed Imai {\it et\,al.}\ to estimate
$2\pi\rho_{s}\approx1730$\,K.\cite{imai93,imai93-2}  Since
$2\pi\rho_{s}=1.13J$, this implies $J\approx1500$\,K, in  good agreement
with estimation based on  neutron and Raman scattering
measurements.\cite{hayden91,singh-raman}

Naturally, one would expect that
measurements of $^{63}1/T_{1}$ may allow us to determine the spin 
stiffness also in the striped phase below
$T_{charge}$.\cite{tranquada4,kim}  Unfortunately, $^{63}1/T_{1}$ measured
for the {\it observable} $^{63}$Cu NQR signal represents the spin
fluctuation properties of the not yet stripe ordered segments (i.e.\
$\omega/T$ scaling), and exhibits little signature of exponential
divergence even below 50 K\cite{singer,suh} where  spin freezing is
observed by neutron scattering.  The absence of divergence in the measured
value of $^{63}1/T_{1}$ can be understood based on the following
considerations.  First, we recall that, quite generally, nuclear
relaxation rates are inversely  proportional to the energy scales of spin
fluctuations, such as
$2\pi\rho_{s}$ and
$J$.\cite{moriya56} Since
$^{63}1/T_{1}$ in hole doped high $T_{c}$ cuprates, including the striped
phase, exhibit nearly identical values with undoped 
La$_{2}$CuO$_{4}$ at higher temperatures,\cite{imai93} the
fundamental energy scale of $^{63}1/T_{1}$ is set  by the same bare $J$
at high temperatures, even in the striped phase.  On the other hand, from
the measurements of the spin-spin  correlation length, Tranquada {\it
et\,al.}\ showed that the effective  spin stiffness $2\pi\rho_{s}^{eff}$
is as small as $\sim$200\,K below $T_{charge}$.\cite{tranquada4} We note 
that the fairly small spin stiffness can be considered a natural 
consequence of slowed charge dynamics, because a hole 
intervening the Cu-Cu exchange path would locally diminish the value of 
exchange interaction $J$, and the effective  spin stiffness
$2\pi\rho_{s}^{eff}$ is essentially a spatial average of $J$ at longer
length scales.
 
We now recall that all of the $^{63}$Cu NQR signal in high $T_{c}$ 
cuprates is near the detection limit of pulsed NQR experiments, and only
a factor $\sim$3 enhancement in $^{63}1/T_{1}$ and an accompanying change
in $^{63}1/T_2$ is sufficient to make signal detection
impossible.\cite{imai93,imai93-2} That is why
$^{63}$Cu NQR measurements in  La$_{2}$CuO$_{4}$ could not be conducted
below $\sim$400\,K.\cite{imai93}  The drastically reduced value of the
spin stiffness in the charge ordered segments  mean that  once 
the charge order sets in, $^{63}1/T_{1}$ and $^{63}1/T_{2}$ in those 
segments begin to blow up inversely proportional to
$2\pi\rho_{s}^{eff}$.  The fact that we do not observe a divergence of
$^{63}1/T_{1}$ {\it for the observable parts of the NQR signal} near 
$T_{charge}$ indicates that this divergence of  $^{63}1/T_{1}$ in 
the charge ordered segments takes place very quickly.  In fact, as
summarized in Fig.~\ref{fig_anchor_points}, the Cu spin fluctuation
frequency decreases by $\sim$\,5-7 orders of magnitude between 70\,K and
350\,mK, roughly an order of magnitude every 10\,K.  This provides a
natural explanation of why the nuclear resonance  signal from those
segments experience dramatic and sudden wipeout due to the extremely fast 
relaxation times.  This is the indirect spin contribution to wipeout 
triggered by charge order.\cite{hunt,singer}

Even though the wipeout of $^{63}$Cu NQR prevents us from observing the
renormalized classical  behavior in striped phase, one can utilize
$^{139}$La to probe the  slowing of spin fluctuations.  The hyperfine
coupling between the Cu electron spins and $^{139}$La nuclear spins is 
two orders of magnitude smaller than for the $^{63}$Cu site.  Since
$1/T_{1}$ is proportional to  the square of the hyperfine coupling, this
means that, if we ignore nonmagnetic contributions to the relaxation,
$^{139}1/T_1$ at the La sites is four orders of magnitude less sensitive
to the divergence caused by the slowing Cu spin dynamics as compared with
$^{63}1/T_1$ at the Cu site.
\begin{equation}
\frac{1}{^{139}T_1}\approx\left(
\frac{^{139}\gamma_n}{^{63}\gamma_n}
\frac{^{139}H_{hf}}{^{63}H_{hf}}
\right)^2
\frac{1}{^{63}T_1}\approx(0.4\times10^{-4})\frac{1}{^{63}T_1}
\label{eq_T1_ratio}
\end{equation}
Thus $^{139}$La NQR signal
does not suffer signal wipeout until the low frequency spin fluctuations
have slowed four more orders beyond the onset of $^{63}$Cu wipeout. 
Accordingly, $^{139}$La NQR $^{139}1/T_1$ is a better measure of spin
fluctuations in the stripe ordered segments of the CuO$_2$ planes below
$T_{charge}$.  However, we caution that $^{139}1/T_1$
is known to be dominated by processes other than Cu spin fluctuations
above $T\sim T_{charge}$, most likely electric quadrupole coupling with
lattice vibrations, making interpretation of $^{139}1/T_1$ above
$T_{charge}$ difficult.

In Fig.~\ref{fig_combo_T1} we plot both $^{63}1/T_{1}$ and~$^{139}1/T_{1}$
for La$_{1.68}$Eu$_{0.2}$Sr$_{0.12}$CuO$_{4}$. As initially discovered by
Chou, Cho, Borsa and co-workers,\cite{cho,chou} $^{139}1/T_{1}$ shows
roughly exponential divergence below $T_{NQR}$ in the underdoped  regime of
La$_{2-x}$Sr$_{x}$CuO$_{4}$.  At $x=0$
the $^{63}$Cu data fits well to the renormalized classical form in the
low $T$ limit ($T\lesssim2\pi\rho_s/2$), written
as\cite{charravarty-orbach}
\begin{eqnarray}
\frac{1}{T_1}=
\frac{0.35}{Z_c}
\frac{A_{\bf Q}^2}{J\hbar}
\frac{\xi}{a}
\frac{(T/2\pi\rho_s)^{3/2}}{(1+T/2\pi\rho_s)^{2}}
\label{eq_RC_T1_xi}\\
\xi=0.27\frac{\hbar c}{2\pi\rho_s}
\frac{\exp(2\pi\rho_s/T)}{1+T/2\pi\rho_s}
\label{eq_RC_correlation}
\end{eqnarray}
with the ${\bf Q}=(\pi,\pi)$ component of the hyperfine coupling
constant $A_{\bf Q}=A_x-4B$, the constant $Z_c=1.18$, and
$2\pi\rho_s=1730$\,K.  The spin wave velocity is
$c=\sqrt{2}JaZ_c/\hbar$. We also found similar behavior of
$^{139}1/T_{1}$ in the striped phase of
La$_{1.8-x}$Eu$_{0.2}$Sr$_{x}$CuO$_{4}$ and  La$_{2-x}$Ba$_{x}$CuO$_{4}$
below $T_{charge}$, where the value of $2\pi\rho_s^{eff}$ is greatly
reduced.  In fact, from Fig.~\ref{fig_combo_T1} we can estimate
$2\pi\rho_s^{eff}\approx200$\,K, which is in good agreement with the
previously mentioned result of neutron scattering,
$2\pi\rho_s^{eff}=200\pm50$\,K.\cite{tranquada4}  It is important to
notice that the extrapolation of the low temperature renormalized
classical fit to higher temperatures, as shown by the dashed curves in
Fig.~\ref{fig_combo_T1}, overestimates $1/T_1$ observed for
$^{63}$Cu NQR by an order of magnitude above $T_{charge}$. 

\subsection{Spin fluctuation spectrum deduced from analysis of $\bf T_1$
recovery data}
\label{subsec_T1}

\begin{figure}[ht] 
\begin{center}
\epsfig{figure=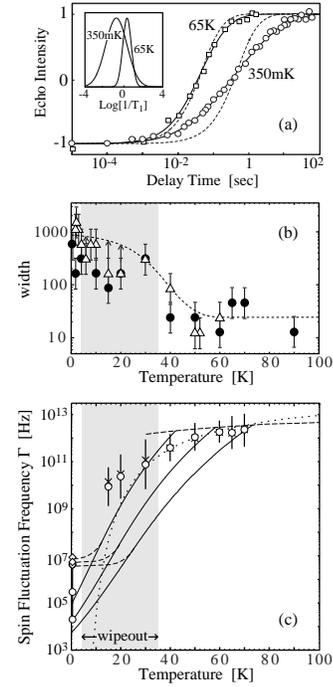,width=2.04in,angle=0}
\caption{(a) $^{139}$La NQR recovery data at 350\,mK and
65\,K for La$_{1.68}$Eu$_{0.20}$Sr$_{0.12}$CuO$_4$ fit to single values of
$T_1$ (dashed curves) and to distributions of $T_1$ (solid curves) as
given in the inset.\cite{note-log-scales}  The width $w$ of the
distributions (defined as the quotient of the two $T_1$ values at the
half maximum positions of the distributions) are plotted as a function of
temperature in (b) for La$_{1.875}$Ba$_{0.125}$CuO$_4$ ($\triangle$) and
La$_{1.68}$Eu$_{0.20}$Sr$_{0.12}$CuO$_4$ ($\bullet$).  We note that the
width is underestimated during NQR signal wipeout (the shaded region) and
with this in mind we draw small vertical arrows from the most affected
measurements  and a dotted curve as a guide for the eye.  In (c) the
fluctuation frequency $\Gamma$ deduced from $T_1$ (from forced fit with
single $T_1$, ($\times$); central value of best fit distribution ($\circ$)
with FWHM as given by the vertical bars).  At 350\,mK we display data
from a Lorentzian extension ($\circ$) of the form of $T_1$ and from a
Gaussian extension ($\diamond$) ({central value and FWHM positions marked
with matching symbols for clarity}).  The
solid and dashed curves are reproduced from Fig.~\ref{fig_anchor_points}b
and act as guides for the eye and the dotted curve represents the form
$\exp(-2\pi\rho_s^{eff}/T)$ with $2\pi\rho_s^{eff}=200$\,K.}
\label{fig_T1_recovery}
\end{center}
\end{figure}

The renormalized classical expression for the nuclear spin-lattice
relaxation rate $1/T_1$ (Eq.~\ref{eq_RC_T1_xi}) can be rewritten in terms
of a frequency scale $\Gamma=(c/\xi)\sqrt{T/2\pi\rho_s}$ as
\begin{equation}
\frac{1}{T_1}=0.49\,
\frac{A_{\bf Q}^2}{\hbar^2}\,
\frac{1}{\Gamma}\,
\frac{(T/2\pi\rho_s)^2}{(1+T/2\pi\rho_s)^2}
\label{eq_RC_T1_Gamma}
\end{equation}
which connects $\Gamma$ to the observable quantity $T_1$, thus
providing a natural route to obtain information on the fluctuation
spectrum.  Due to the strong wipeout of the $^{63}$Cu NQR signal, we are
forced to rely on the $^{139}$La relaxation data, where wipeout is less
severe and limited to the range 5 - 35\,K (see
Fig.~\ref{fig_wipeout_fits} for wipeout data).   We measure $^{139}1/T_1$
by the inversion-recovery method.  The recovery of the nuclear
magnetization may be fit to the solution to the standard rate equations,
\begin{equation}
\frac{m(t)}{m(\infty)}=1
-\frac{3}{7}e^{-\frac{3t}{T_1}}
-\frac{100}{77}e^{-\frac{10t}{T_1}}
-\frac{3}{11}e^{-\frac{21t}{T_1}}
\label{eq_T1_recovery}
\end{equation}
for the $|\pm\frac{5}{2}\rangle \leftrightarrow 
|\pm\frac{7}{2}\rangle$ NQR transition where $m(t)$ is the magnetization
at a time $t$ and $1/T_1$ is the single relaxation rate.  This form is
valid for magnetic relaxation when the spectral function is constant
across the three NQR transitions, which is generally the case for
$\Gamma\gg f_{NQR}$.  At 65\,K, the fit of the recovery data with a single
value of $T_1$ is reasonably good, as seen in Fig.~\ref{fig_T1_recovery}a,
but at 350\,mK, the fit is quite poor.  This led us to simulate the
recovery curves due to distributions of $T_1$.  The distributions in
$1/T_1$ are displayed in the inset of  Fig.~\ref{fig_T1_recovery}a, and
the recovery is calculated by summing contributions from each segment of
the sample, where the magnetization of each individual segment recovers
per Eq.~\ref{eq_T1_recovery}.  These simulated recovery curves are
plotted as solid lines in Fig.~\ref{fig_T1_recovery}a and provide a much
better fit to the 350\,mK data, while slightly improving the quality of
the fit at 65\,K.  The small distribution evident within the 65\,K
recovery data does not necessarily reflect an intrinsic distribution of the
magnetic relaxation time $T_1$ but may be due to
extrinsic processes such as quadrupole relaxation.  In
general, the central value of the $T_1$ distribution is very close to that
which we find by force fitting to a single relaxation rate.  As seen in
Fig.~\ref{fig_T1_recovery}b, we found that the width $w$ of the $T_1$
distribution increases with decreasing temperature below $T_{charge}$ from
$w$\,$\sim$\,20 to $w$\,$\sim$\,200-1000 at 1.7\,K and below.  The local
minimum near 15\,K is caused by the fact that $^{139}$La NQR signals with
extremely fast $^{139}1/T_1$ are effectively wiped out due to short $T_2$
relaxation rates (see Fig.~\ref{fig_La_Eu_wipeout_fraction}), hence the
measured nuclear spin recovery does not include those contributions.  In
the absence of those contributions, the width $w$\,$\sim$\,100 near 15\,K
should be considered as a lower bound on $w$.

Using the RC form (Eq.~\ref{eq_RC_T1_Gamma}) with
$2\pi\rho_s^{eff}=200$\,K we can convert these distributions of
$1/T_1$ to $\Gamma$ without any additional parameters.   For $\Gamma\gg
f_{NQR}$ the conversion is straightforward; these points are shown in
Fig.~\ref{fig_T1_recovery}c.  For $\Gamma$ close to $f_{NQR}$ the
renormalized classical form may be rewritten as
\begin{equation}
\frac{1}{T_1}=0.49\,
\frac{A_{\bf Q}^2}{\hbar^2}\,
\frac{1}{\Gamma}\,
\frac{1}{(1+f_{NQR}^2/\Gamma^2)}
\frac{(T/2\pi\rho_s)^2}{(1+T/2\pi\rho_s)^2}, 
\label{eq_RC_lorentzian}
\end{equation}
where $f_{NQR}=18$\,MHz and we assume that the spin
dynamics follow the Lorentz form in the renormalized classical regime. 
This function is now double valued in $\Gamma$, which, in addition to
signal wipeout, makes the determination of $\Gamma$ in the region 5-35\,K
difficult.  In that region, $\Gamma\sim f_{NQR}$, which implies fast
relaxation and signal wipeout.  From $T_1$ measurements alone it is
impossible to know how much of the distribution of $\Gamma$ is greater or
less than $f_{NQR}$.  All we can say for certain is that experimentally
the value of $1/T_1$ is underestimated because the fastest sections are
not observable.  However, at 
350\,mK these difficulties disappear because all of the signal intensity
is recovered and we know that $\Gamma<f_{NQR}$.  We can now safely employ
Eq.~\ref{eq_RC_lorentzian} to determine the distribution of spin
fluctuation frequencies, which is centered at $3\times10^5$\,Hz with a
width of more than two orders of magnitude.

In the preceeding analysis for $\Gamma\lesssim f_{NQR}$ we chose to extend
the RC form of $1/T_1$ with a factor of $1/(1+f_{NQR}^2/\Gamma^2)$
(Eq.~\ref{eq_RC_lorentzian}).  This assumption is certaintly justified
for the paramagnetic spin dynamics above 70\,K where the neutron 
data\cite{aeppli} fits  well to a Lorentzian.  However, there is no
reason to believe  that the very slow dynamics in spin stripes at 350\,mK
have to be governed  by the same type of relaxational mechanism.  For
instance, if scattering by  meandering hole rivers takes over as the
dominant mechanism of spin  fluctuations at low temperatures, 
the form of the relaxation function may change.    Let us try another
sensible choice, the form based on Moriya's Gaussian 
approximation,\cite{moriya-gaussian} which would give frequency
dependence in $1/T_1$ as
\begin{equation}\frac{1}{T_1}=0.49\,
\frac{A_{\bf Q}^2}{\hbar^2}\,
\frac{1}{\Gamma}\exp
\left[-\frac{f_{NQR}^2}{2\Gamma^2}\right]
\frac{(T/2\pi\rho_s)^2}{(1+T/2\pi\rho_s)^2}.
\label{eq_RC_gaussian}
\end{equation}
We also employ the form to deduce $\Gamma$ at 350\,mK, and find a much
narrower distribution centered at a higher frequency,
$5\times10^6$\,Hz.  We will discuss the implications of
choosing either the Lorentzian or Gaussian from of $1/T_1$ in section
\ref{subsec_slowing_summary}.  We note that in simulating the $T_1$
recovery curve we continue to employ Eq.~\ref{eq_T1_recovery} even for
$\Gamma$ close to and less than $f_{NQR}$.  In that region the spectral
function is not flat over the three resonant transitions, invalidating the
basis of Eq.~\ref{eq_T1_recovery}.   Continued use of
Eq.~\ref{eq_T1_recovery} is justifiable because the distribution in
$^{139}1/T_1$ extends over several orders of magnitude and should be
much  greater than the potential error induced through the continued use
of Eq.~\ref{eq_T1_recovery}.

\subsection{Numerical simulations of NQR signal wipeout
and recovery, neutron, and $\mu$SR}
\label{subsec_wipeout}


As previously defined, wipeout is the amomalous loss of NQR signal
intensity.  For La$_{2-x-y}$(Nd,Eu)$_y$Sr$_x$CuO$_4$ with
$0.12\approx\frac{1}{8}$, the onset temperature $T_{NQR}$\cite{T-NQR} of
the $^{63}$Cu NQR wipeout agrees very well with the onset temperature
$T_{charge}$  of short-range charge order detected by neutron
\cite{tranquada1,tranquada2,tranquada3,tranquada4,ichikawa} and x-ray 
scattering measurements.\cite{zimmermann,niemoller}  Moreover, the
fraction of wipeout $F(T)$ shows  a characteristic temperature dependence
which is nearly identical to the charge order  parameter observed by
scattering techniques (i.e.\ the square root of the scattering intensity
of the charge order peak), leading us to argue that the  temperature
dependence of the wipeout fraction $F(T)$ is {\it a good 
measure}\cite{hunt} of the charge order parameter. Recent  $^{139}$La
NMR studies of La$_{2}$NiO$_{4+\delta}$ by Abu-Shiekah {\it
et\,al.}\cite{abu}  also showed that the temperature dependence of the 
wipeout fraction of
$^{139}$La NMR in a high magnetic field (with an onset temperature of
$\sim$\,200\,K)  agrees quite well with the charge stripe order
parameter\cite{Ni-charge} observed by elastic neutron scattering  by
Tranquada, Wochner and coworkers.\cite{tranquada7,Wochner}

Naively thinking, the slowing charge fluctuations can 
make {\it direct} contributions to wipeout through extreme NQR line 
broadening similar to the case of conventional CDW 
systems,\cite{butaud,ross,nomura} and also through extremely fast
quadrupole relaxation  times.\cite{abragam,slichter}  It is difficult to
access how strongly the charge fluctuations near $T_{charge}$ affect the
NQR properties of the missing signals because the signals from the
already stripe ordered segments are not observable, similar to the {\it
all or nothing} behavior of wipeout observed for
Cu:Mn.\cite{MacLaughlin-Alloul}

In addition to the direct charge contribution to wipeout, we have also
pointed out that {\it  charge order  turns on low frequency spin
fluctuations}  below $T_{NQR}$ within the charge ordered 
domains, which in turn causes wipeout by making $^{63}T_{1}$ and 
$^{63}T_{2}$ too short to detect the $^{63}$Cu NQR spin echo
signal.\cite{hunt}  That is, even if charge fluctuations are not strong
enough to cause the wipeout of $^{63}$Cu NQR directly, charge order
creates patches of the CuO$_2$ plane that behave similarly to conventional
spin glasses at slow time scales.  As already noted by Hunt {\it
et.\,al.}, dilute magnetic moments in metals (e.g.\ Cu:Fe\cite{nagasawa})
are known to cause wipeout of the NMR signal of the host metal.  Restating
this, we can say that the slowing charge dynamics are the {\it indirect}
cause of $^{63}$Cu NQR wipeout by triggering the drastic slowing of spin
fluctuations within the regions where holes are trying to form charge
rivers.  Unfortunately, these important statements from Hunt {\it
et.\,al.}\ and Singer {\it et.\,al}\ regarding the effects of strong spin
dynamics on wipeout have been overlooked by some readers, including
the authors of a recent publication from Los Alamos.\cite{curro}

Hunt {\it et\,al.}\ cited several pieces of experimental evidence that
are consistent with the picture that anomalous slowing of spin dynamics
are caused by slowing charge dynamics.   First, inelastic neutron
scattering measurements conducted  at low energy transfers
(1.75-3.5\,meV) seem to show an anomalous enhancement of the low
frequency spin fluctuations near and  below $T_{charge}$, even before
quasi-elastic neutron scattering with  energy transfer less than 1\,meV
detects  frozen spins.\cite{tranquada4}

Second, the spin-lattice relaxation rate $^{139}1/T_{1}$ at the
$^{139}$La sites  begins to exhibit an anomalous upturn below $T_{NQR}$.
This enhancement of $^{139}1/T_{1}$ is followed by a roughly
exponential divergence at lower temperatures, $^{139}1/T_{1}\sim 
\exp(2\pi\rho_{s}^{eff}/k_{B}T)$, where $2\pi\rho_{s}^{eff}$ is the
 effective spin  stiffness which is of the order $10^{1}$-$10^{2}$\,K, 
as already shown by earlier comprehensive $^{139}$La NQR measurements by Chou, 
Cho, Borsa, Johnston and co-workers in underdoped
La$_{2-x}$Sr$_x$CuO$_4$.\cite{cho,chou}   Recalling that the
exponential divergence  of $1/T_{1}$ is a typical signature of the low
temperature  {\it renormalized classical scaling} behavior of 
$S=\frac{1}{2}$ quantum Heisenberg systems in both 
isotropic\cite{borsa,imai93} and  anisotropic 2d
systems,\cite{thurber-3leg} the exponential divergence of 
$^{139}1/T_{1}$ below $T_{NQR}$ is consistent with a rather drastic reduction of the 
effective spin stiffness $\rho_{s}^{eff}$ caused by slowing charge 
dynamics.  The spin-spin correlation length measured by Tranquada {\it
et\,al.}\ also supports the renormalized, small spin stiffness with 
$2\pi\rho_{s}^{eff}\approx200$\,K.\cite{tranquada4}   In contrast, we
recall that the  fundamental energy scale in hole doped
La$_{2-x}$Sr$_x$CuO$_4$ is set  by $2\pi\rho_s\approx1730$\,K as shown by
NQR\cite{imai93,imai93-2} and neutron  scattering.\cite{hayden}   High
energy scales of the order of 600\,K are also inferred from a scaling
analysis of the uniform spin susceptibility by Johnston\cite{johnston}
and Hwang {\it et\,al.}.\cite{hwang} This suggests that drastic reduction
in spin  fluctuation energy scales from the order of 10$^3$\,K to
10$^2$\,K occurs near the charge ordering temperature.
This has an important consequence on  the wipeout phenomenon, because
 quite generally the overall magnitude of nuclear relaxation rates 
$^{63}1/T_{1}$ and $^{63}1/T_{2}$ are inversely proportional to 
the spin fluctuation energy scale
$^{63}1/T_{1,2}\propto1/2\pi\rho_s^{eff}$.\cite{moriya56}   The observed
relaxation rate $^{63}1/T_{1}\sim2000$\,sec$^{-1}$ is quite fast even
above  $T_{charge}$ in hole-doped high $T_{c}$ cuprates, while
$^{63}$Cu NQR signals  become unobservable when relaxation rates exceed
$^{63}1/T_{1}\sim6000$\,sec$^{-1}$ for measurements with the
quantization axis along the crystalline c-axis.\cite{imai93} 
When $^{63}1/T_1$ exceeds $\sim6000$\,sec$^{-1}$ the relaxation times
$^{63}T_1$ and $^{63}T_2$ become shorter than the limit of experimentally
accessible times.  In other words, the $^{63}$Cu NQR signal in {\it all}
high $T_c$ cuprates is on the verge of wipeout due to strong low frequency
spin fluctuations. This means that a factor 2$\sim$3 reduction in spin 
stiffness $\rho_{s}^{eff}$ near $T_{charge}$ caused by slowed charge
dynamics would immediately wipe out $^{63}$Cu NQR signals.  This explains
why the $^{63}$Cu NQR wipeout fraction accurately tracks the charge order
parameter as measured by scattering experiments if one interprets the
latter as a measure of the volume fraction of charge stripe ordered
segments.

Third, the Gaussian component of spin-spin relaxation process 
$^{63,65}1/T_{2}$ at the $^{63,65}$Cu sites crosses over to
Lorentzian  (single-exponential) decay below $T_{charge}$ by motional
narrowing effects  due to fluctuating magnetic hyperfine 
fields.\cite{hunt,singer}  The dominance of spin
rather than charge fluctuations to the mechanism of  the Lorentzian spin
echo decay was suggested by the isotope ratio between
$^{63,65}$Cu.\cite{hunt,singer}   This crossover of spin echo decay,
accompanied by NQR/NMR intensity  wipeout, is known to be a typical 
signature of slowing down of conventional spin-glass systems 
such as Cu:Mn.\cite{MacLaughlin-Alloul,chen}
Thus all of these results as well as the dramatic increase in the EPR
linewidth\cite{kataev} are consistent with a simple picture that
spin-glass like behavior sets in when charge dynamics slow down, and the
combined effects of slowing charge and spin stripe fluctuations can 
naturally provide a qualitative account of the wipeout behavior for 
$x\gtrsim\frac{1}{8}$ below $T_{NQR}\approx T_{charge}$.


In a recent article, Curro, Hammel and coworkers\cite{curro} made an important
observation that enabled quantitative estimation of the {\it indirect}
spin contribution to wipeout.\cite{hunt,singer}  The essence of the
analysis of Curro and Hammel is as follows.   First, instead of saying
that {\it extremely short $^{63}T_{1}$ and $^{63}T_{2}$ result in
wipeout},\cite{hunt,singer} they note that {\it the $^{63}C$u NQR signal
cannot be detected if $^{63}T_{1}$ and $^{63}T_{2}$ are shorter than
certain cutoff values labeled $^{63}T_{1}^{critical}$ and
$^{63}T_{2}^{critical}$}.   Using several assumptions, they estimate
these cutoff values.  Once $^{63}T_{1}^{critical}$ is  deduced, Curro {\it
et\,al.}\ use the BPP form\cite{slichter} of $T_1$ to determine the
corresponding spin fluctuation frequency $^{63}\Gamma^{wipeout}$.  The
wipeout fraction is the weight of the spin fluctuation distribution that
is  slower than $^{63}\Gamma^{wipeout}$ because the slower segments have
$^{63}T_{1}$ and $^{63}T_{2}$ that are too fast to allow the detection of
the spin echo.

Inspired by their analysis, we attempt to fit not only the wipeout of the
$^{63}$Cu NQR signal below $T_{NQR}\approx70$\,K but also its recovery below
8\,K, the wipeout of $^{139}$La NQR signal below 40\,K and its recovery
below 18\,K, the spin order  parameter measured by elastic neutron
scattering below 50\,K and the $\mu$SR  asymetry below 30\,K by taking
into account the qualitative change in spin dynamics below $T_{charge}$. 

We take the following approach, which relies primarily
on experimental data instead of theoretical  assumptions.  In order to
set $^{63}\Gamma^{cutoff}$ we note that
$^{63}1/T_1\approx2000\,\mbox{sec}^{-1}$ at 70\,K, at which temperature we
know that the energy scale of spin fluctuations is $h\Gamma\approx15$\,meV
from neutron scattering. From earlier $^{63}$Cu NQR studies in
paramagnetic  La$_{2}$CuO$_{4}$,\cite{imai93} we empirically know that
the loss of $^{63}$Cu NQR intensity due to fast relaxation occurs when
the relaxation  rate
exceeds $^{63}1/T_1^{critical}\approx6000\,\mbox{sec}^{-1}$.  Following
Eq.~\ref{eq_RC_T1_Gamma}, this implies that 
$^{63}\Gamma^{wipeout}$ is approximately a factor of three lower than the
value of $h\Gamma\approx15$\,meV at 70\,K.  In this way, we define
$^{63}\Gamma^{wipeout}\equiv5$\,meV, which is 
$1.2\times10^{12}$\,Hz (so
that $1/^{63}\Gamma^{wipeout}\approx0.83$\,psec).  

The cutoff for the wipeout of $^{139}$La NQR signal is found in an analogous
manner.  From our experments shown below in
Fig.~\ref{fig_La_Eu_wipeout_fraction} and Fig.~\ref{fig_1/T1T}, we
see that $^{139}1/T_1$ rises to approximately $80\,\mbox{sec}^{-1}$ when
wipeout is most severe.  The remaining portion of the sample is on the
verge of wiping out and has $T_1$ very close to $^{139}1/T_1^{critical}$. 
Thus we take  $^{139}1/T_1^{critical}\equiv80\,\mbox{sec}^{-1}$.  We now
need to convert this value of $^{139}1/T_1^{critical}$ to the
corresponding frequency scale.  Using the renormalized classical form
(Eq.~\ref{eq_RC_T1_Gamma}) in the high frequency limit 
($\Gamma\gg\omega_n$) we see that
\begin{equation}
\frac{1}{T_1}\propto\frac{(\gamma_n
H_{hf})^2}{\Gamma}\frac{(T/2\pi\rho_s)^2}{(1+T/2\pi\rho_s)^2},
\label{eq_BPP_high_freq}
\end{equation}
where wipeout occurs at 70\,K for $^{63}$Cu and at 40\,K for $^{139}$La, 
and we take $2\pi\rho_s\approx200\,$K as determined above.  Since we know
the ratio of the gyromagnetic ratios to be
$^{139}\gamma_n/^{63}\gamma_n= 0.533$ and the ratio of the hyperfine fields
at the two nuclear positions to be
$^{139}H_{hf}/^{63}H_{hf}= 0.012$ as inferred
from Zeeman perturbed NQR in La$_2$CuO$_4$,\cite{kitaoka,tsuda}  we
immediately calculate $^{139}\Gamma^{wipeout}=5.1\times10^8$\,Hz.  We note
that the $^{139}F(T)$ data is taken with a fixed delay time
$\tau=40\,\mu$sec, which is justified due to the slow relaxation rate
$^{139}1/T_2$ at the La site (see the inset of
Fig.~\ref{fig_La_Eu_wipeout_fraction}).

At low temperatures, the $^{63}$Cu and $^{139}$La NQR signal that was
wiped out, begins to reemerge.  The return of signal begins at about 8\,K
for $^{63}$Cu and 18\,K for $^{139}$La.  Within our model, the
reemergence of signal is not surprising, in fact, it is to be expected if
the frequency range falls low enough so that $T_1$ and $T_2$ become
longer.  Using the Lorentzian extension of the renormalized classical
form (Eq.~\ref{eq_RC_lorentzian}), with the NQR resonance frequency
$f_{NQR}$ taken as 6\,MHz and 35\,MHz for $^{139}$La and $^{63}$Cu,
respectively, that  the relaxation rate $1/T_1$ achieves its
maximum value when the characteristic fluctuation frequency $\Gamma$
coincides with  $f_{NQR}$.  Below that, additional slowing of the
fluctuations (decreasing of $\Gamma$) reduces the relaxation rate
$1/T_1$, so that at some low frequency scale, which is defined as
$\Gamma^{recovery}$, the following  condition is satisfied:
$1/T_1(\Gamma^{recovery})=1/T_1(\Gamma^{wipeout})=1/T_1^{critical}$
where we include temperature factors from Eq.~\ref{eq_BPP_high_freq}.  We
note that $\Gamma^{recovery}<f_n<\Gamma^{wipeout}$.  Within this model, it
is straightforward to deduce that
that $^{139}\Gamma^{recovery}=2.9\times10^5$\,Hz and
$^{63}\Gamma^{recovery}=4.0\times10^4$\,Hz.  This means that the slowest
components within that sample slow to $2.9\times10^5$\,Hz at approximately
18\,K, providing $^{139}$La NQR signal recovery, and continue to slow
through $4.0\times10^4$\,Hz at 8\,K where they are manifested as
recovered $^{63}$Cu NQR signal.

\begin{figure}[ht]
\begin{center}
\epsfig{figure=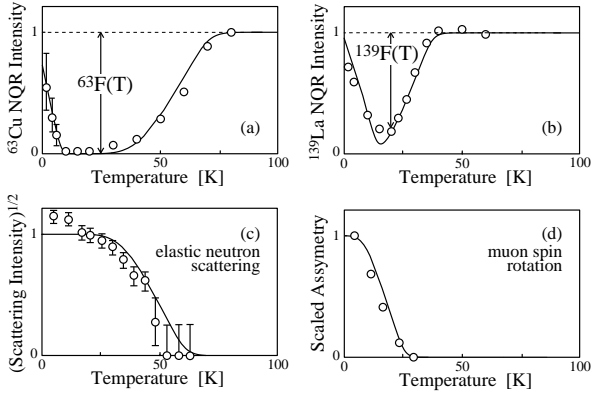,width=3.2in,angle=0}
\end{center}
\caption{Four sets of experimental data and
corresponding numerical simulations based on the model
described in the text, using the $D(T,f)$ shown in
Fig.~\ref{fig_anchor_points}b. (a) $^{63}$Cu wipeout in
La$_{1.68}$Eu$_{0.2}$Sr$_{0.12}$CuO$_{4}$, (b) the $^{139}$La wipeout
from the $|\pm\frac{1}{2}\rangle\leftrightarrow |\pm\frac{3}{2}\rangle$
transition in La$_{1.68}$Eu$_{0.2}$Sr$_{0.12}$CuO$_{4}$ 
(The  wipeout behavior at $x\approx \frac{1}{8}$ is nearly identical for
Eu, Nd, and  Ba co-doped samples.) (c) the spin order parameter in
La$_{1.48}$Nd$_{0.4}$Sr$_{0.12}$CuO$_{4}$ deduced from elastic neutron
scattering as the square root of the scattered intensity, from Tranquada
{\it et\,al.},\cite{tranquada1} and (d) the $\mu$SR asymetry factor
$A_\perp/A_{total}\times\nu^2$ in
La$_{1.475}$Nd$_{0.4}$Sr$_{0.125}$CuO$_{4}$,
from Nachumi {\it et\,al.}.\cite{nachumi}}
\label{fig_wipeout_fits}
\end{figure}

Combining all of these pieces of information regarding the spin fluctuation 
time scales in the stripe phase, we construct a chart showing the 
anchoring points of the temperature dependence of spin fluctuation
frequency scale $\Gamma$, as  presented in Fig.~\ref{fig_anchor_points}b. 
Using these anchoring points, we smoothly interpolate to define a single
distribution function $D(T,f)$, which gives the relative fraction of the
sample volume that experiences fluctuations at frequency $f$ at temperature
$T$.  This distribution is depicted by the black and gray contours in
Fig.~\ref{fig_anchor_points}.  We begin with an educated guess at the
form of distribution function, and refine the function as we iteratively
produce a better and better fit to the experimental data as presented
throughout this section.  In this way we deduce the actual fluctuation
spectrum based on the experimental data.  Since there are no experimental
constraints for $\Gamma$ in the range above $^{63}\Gamma^{wipeout}$ but
less than the $\omega/T$ scaling line, we are forced to extrapolate in
that region.  In Fig.~\ref{fig_anchor_points} we depict smooth
extrapolations of $D(T,f)$ that terminate at the $\omega/T$ scaling line
at a temperature above $T_{charge}$.  However, we note that we are
ignoring the possible contribution of the slowing of charge dynamics to
Cu NQR wipeout effects.

The simulated wipeout fraction $F(T)$ is the fraction of the NQR signal
that is not observable,  which corresponds to the portion which
experiences fluctuations that are slower than $\Gamma^{wipeout}$ but
faster than $\Gamma^{recovery}$.  This can be written as,
\begin{equation}
F(T)=\frac{\int_{\Gamma^{recovery}}^{\Gamma^{wipeout}}D(T,f) df}
{\int_{0}^{\infty}D(T,f) df}
\label{eq_wipeout_fraction}
\end{equation}
where the denominator acts as a normalization condition on $D(T,f)$ at every
temperature.  For the neutron scattering and $\mu$SR measurements, the
fraction of the sample that has ordered is given by a similar form,
\begin{equation}
G^{probe}(T)=\frac{\int_{0}^{\Gamma^{probe}}D(T,f)
df} {\int_{0}^{\infty}D(T,f) df}
\label{eq_neutron_muSR}
\end{equation}
where $probe$ is either $\mu SR$ or $neutron$.

All four of these quantities ($^{63}F(T)$, $^{139}F(T)$, $G^{neutron}(T)$
and $G^{\mu SR}(T)$) have been numerically computed over a range of
temperatures using the distribution $D(T,f)$ presented in
Fig.~\ref{fig_anchor_points}b, and the resulting curves are plotted in
Fig.~\ref{fig_wipeout_fits} along with the relevant experimental data. 
The simulations work remarkably well in reproducing both the $^{63}$Cu and
$^{139}$La wipeout data, as well as the neutron and $\mu$SR results.  As
mentioned previously, the quality of these fits is the result of iterative
alterations of trial distribution functions that eventually led us to 
the distribution $D(T,f)$ shown in Fig.~\ref{fig_anchor_points}b.
We find that a distribution function based on a Gaussian in $\log f$ with
different widths to higher and lower frequency works well.  
However, the use of an iterative method means that we cannot exclude the
possibility that a somewhat different $f$ and $T$ dependence of
$D(T,f)$ may work as well as this one,  although it is difficult to
imagine that a qualitatively different form would succeed, given the large
number of experimental constraints that must be satisfied.

Although we do not explicitly utilize the nuclear spin lattice relaxation
rate $1/T_1$ in the analysis of this section, our choice of the
form of $T_1$ for $\Gamma$ below the Larmor frequency does play an
important role in setting the vertical position of the anchoring points in
Fig.~\ref{fig_anchor_points}.  To this point we have utilized the
Lorentzian form (Eq.~\ref{eq_RC_lorentzian}), but as discussed in section
\ref{subsec_T1}, we might do better to employ the Gaussian form
(Eq.~\ref{eq_RC_gaussian}).  Although this change does not alter any of
the analysis for $\Gamma>5\times10^7$\,Hz, it does substantially change
the behavior for smaller $\Gamma$.  Specifically, the cutoff frequencies
for the recovery of both $^{63}$Cu and $^{139}$La NQR signal are shifted
upwards to $\sim$\,$5\times10^6$\,Hz, causing the distribution $D(T,f)$ to
saturate at low temperatures as shown by the dashed curves in
Fig.~\ref{fig_anchor_points}b and in Fig~\ref{fig_T1_recovery}.  We will
discuss the merits of the Gaussian form of $1/T_1$ in section
\ref{subsec_slowing_summary}.

\begin{figure}
\begin{center}
\epsfig{figure=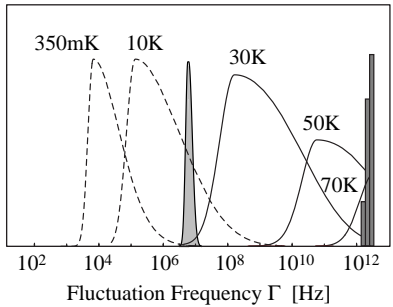,width=2.3in,angle=0}
\end{center}
\caption{Slices of the frequency distribution $D(T,f)$ as deduced from
the analysis of section \ref{subsec_wipeout}.\cite{note-log-scales}  The
shaded rectangles represent the spectral weight in the sharply defined
$\omega/T$ energy scale (70\,K is the right most, 50\,K in the center and
30\,K at the left), which decreases in magnitude at lower temperature as
spectral weight is shifted to the contribution from the stripe ordered
regions represented by the broad curves.  The dashed curves arise from a
Lorentzian extension of the form of $T_1$ (Eq.~\ref{eq_RC_lorentzian}) and
the shaded curve corresponds to a Gaussian extension
(Eq.~\ref{eq_RC_gaussian}) at 350\,mK.}
\label{fig_distribution_slices}
\end{figure}

\subsection{Summary of fluctuation timescales and comparisons to related
works}
\label{subsec_slowing_summary}

Through an analysis of both the $^{139}1/T_1$ recovery data and the
$^{63}$Cu and $^{139}$La wipeout data we arrived at two estimates of
the spin fluctuation spectra as a function of temperature as shown in
Figs.~\ref{fig_anchor_points}b and \ref{fig_T1_recovery}.  As an
alternative presentation of the data in Fig.~\ref{fig_anchor_points}b, we
construct Fig.~\ref{fig_distribution_slices}, which contains slices
through the distribution $D(T,f)$ at selected temperatures.  We find
that there is a sharp change from the high temperature
$\omega/T$ dominated behavior to the much faster exponential slowing of
spin fluctuations below $T_{charge}$.  We understand this qualitative
change in the magnetic fluctuation spectrum to be due to charge ordering. 
The two methods of deducing $\Gamma$ are in qualitative agreement, even
though the quadrupole contribution to relaxation and wipeout makes the
approach based on $^{139}1/T_1$ somewhat questionable.  The differences
between the two methods are smaller than the effects of choosing a
Lorentzian or Gaussian extension of the form of $T_1$.  Without additional
information we would be able to proceed no farther\,---\,however we know
from the lineshape analysis that our 350\,mK lineshape is still
experiencing some motional averaging.  In the Lorentzian schemes more than
50\% of the distribution of $\Gamma$ is lower than
$\Gamma=3\times10^4$\,Hz, which corresponds to 30\,$\mu$sec, the duration
of a pulsed NQR experiment.  This indicates that motional averaging is
unlikely to be effective for the bulk of the sample and we would expect to
see frozen structures in the low temperature Zeeman perturbed spectra,
which are not observed.  However, using a Gaussian model we find that
almost all of the sample experiences fluctuations faster than the
experimental duration, indicating the likelyhood of strong motional
averaging.  Thus, reality is closer to the Gaussian form in order to
preserve motional averaging.  We also note that the low temperature
saturation of $\Gamma$ that naturally arises from the aforementioned
analysis within the Gaussian form is consistent with the saturation of the
spin-spin correlation length observed below 20\,K.\cite{tranquada4}

Before continuing, we point out some qualitative 
similarity between $D(T,f)$ for the present case of the striped cuprates
below $T_{charge}$ and for  the conventional spin glass system Cu:Mn
investigated by Murani by cold  neutron scattering
studies.\cite{murani} It is also important to note that at the slow
timescales of NQR, the striped phase never looks to be striped, but rather
behaves more or less like a conventional spin glass.  This similarity with
the cluster spin glass phase has been repeatedly emphasized by Borsa and
coworkers\cite{cho,chou,borsa} since before the discovery that the spin
glass phase is actually striped by neutron
scattering.\cite{tranquada1,suzuki1998,kimura,wakimoto,lee,tranquada2,tranquada3} 
However, we note that the striped phase does differ from conventional
spin glasses, even at slow timescales.  Specifically, we point out that
the real part of the susceptibility shows only small Curie enhancement,
indicating the scarcity of free spins.  In fact, comparing
La$_{1.87}$Ba$_{0.13}$CuO$_4$\cite{sera} and Zn impurity doped
La$_{1.85}$Sr$_{0.15}$Cu$_{0.98}$Zn$_{0.02}$O$_4$\cite{ikegawa} where 2\%
of the moments are intentionally made to be free, we find that the Curie
component of the susceptibility is nearly two orders of magnitude smaller
in the Ba doped material below $T_{charge}$ than in the Zn doped case. 
This implies that the number of free spins in
La$_{1.87}$Ba$_{0.13}$CuO$_4$ is very small.  It is more likely that the
spin glass like behavior is caused by slow fluctuations of self
organized, short range ordered segments with spin and charge density
modulation.

Recently, Curro, Hammel {\it et\,al.}\cite{curro} and Teitel'baum {\it
et\,al.}\cite{teitelbaum-archive} have reported methods of simulating the
wipeout effects.  In both cases the authors attribute $^{63}$Cu NQR signal
wipeout to slowing spin fluctuations.  We emphasize that the importance of
slowing spin fluctuations on NQR signal wipeout through extremely large
relaxation rates $^{63}1/T_1$ and $^{63}1/T_2$ (the {\it indirect}
contribution), has already been discussed repeatedly in our earlier
publications.\cite{hunt,singer}  The contribution of Curro, Hammel and
coworkers is in introducing of a numerical cutoff to fit wipeout
effects, making quantitative analysis possible.\cite{curro} 
Their pioneering work opened a new path to gain quantitative information
regarding the distribution of the spin fluctuation frequency $\Gamma$. 
This is an important idea that is applicable to the analysis of various
NMR data and they should be fully credited for their contribution.
However, Curro {\it et\,al.}\ overlooked the fact that the importance of
spin fluctuations to wipeout (the {\it indirect} contribution) had
originally been proposed by us, and quoted our publications only in the
context of slowing charge dynamics.  We believe that this
inaccurate quotation is misleading at best.  Even though they have implied
that their way of understanding the wipeout effects was entirely novel, we
must emphasize that the physical processes utilized in their analysis have
little difference from the ideas outlined in Hunt {\it et\,al.}\
regarding the {\it indirect} process, except for one
crucial point.  We believe that the onset of charge
order (or slowing of charge dynamics for $x\lesssim\frac{1}{8}$ due to
random localization effects\cite{singer}) is the cause of the drastic
changes in the magnetic properties that result in NQR signal wipeout.
On the other hand, in the analysis of Curro and coworkers it
is {\it assumed} that glassy slowing of Cu spin fluctuations continuously
proceeds from 300\,K to 4\,K without any qualitative changes in spin
dynamics through
$T_{charge}$.  That is, Curro {\it et\,al.}\ assume that the same
activation temperature dependence of
$\Gamma(T)\propto\exp(-E_a/k_BT)$ properly represents the slowing of Cu
spin fluctuations both above and below $T_{charge}$.  They fit
$^{139}1/T_1$ below 30\,K ($<T_{charge}$) to deduce the activation
energy $E_a$ and its distribution $\Delta$.  Ignoring the drastic
change in the charge degree of freedom near
$T_{charge}$ and the resulting changes in spin
dynamics as reflected in $^{139}1/T_1T$, EPR,\cite{kataev} etc., these
parameters are taken to remain unchanged even above $T_{charge}$. 

We note several inconsistencies arising from the assumptions of the Los
Alamos group.   We point out that the fit of
$^{139}1/T_1$ by Curro and coworkers fails on their own data set above
30\,K.  In addition, within the model of Curro {\it et\,al.},
$^{63}1/T_1$ has the same Gaussian distribution as $\Gamma(T)$ with the
same activation type temperature dependence,
$^{63}1/T_1\propto\exp(E_a/k_BT)$ which is taken to hold for all
temperatures.  Unfortunately, this assumed exponential {\it increase} of
$^{63}1/T_1$ is in sharp  contradiction with the experimental finding that
$^{63}1/T_1$ {\it decreases} with decreasing temperature down to
$T_{charge}$ as shown in Fig.~\ref{fig_combo_T1}. We note that the assumed
exponential form above $T_{charge}$ is essentially equivalent to
extrapolating  $^{63}1/T_1$ along the dotted lines of
Fig.~\ref{fig_combo_T1} between $T_{charge}$ and 300\,K.  Thus, there is
no justification for the assumed forms of $\Gamma(T)$ and $^{63}1/T_1$
above $T_{charge}$.  Since the calculated wipeout fraction by Curro {\it
et\,al.}\ relies on the exponential temperature dependence of $^{63}1/T_1$,
we must conclude that  qualitative success in reproducing wipeout starting
at as high as 150\,K is merely a coincidence that depends strongly on their
choice of $^{63}\Gamma^{wipeout}$.

Finally, although we believe that slowing of charge dynamics is of extreme
importance and that the analysis of the Los
Alamos group is not applicable for $x\gtrsim\frac{1}{8}$, we note that
their assumed form of $\Gamma(T)$ is similar to our finding for $x=0.07$
below
$T_{NQR}$.  However, we note that in this region the beginning of charge
localization has been observed by Ichikawa {\it et\,al.}
at $\sim T_{NQR}$.\cite{ichikawa} (See section
\ref{subsec_underdoped} for a discussion of the underdoped
materials.)  Hence, one may consider the slowing of charge dynamics as the
cause of the glassy slowing of spin fluctuations regardless of spatial
coherence, which is gradually lost with decreasing $x$.\cite{hunt,singer}

After an early version of this work was presented
elsewhere,\cite{imai-houston} the NQR wipeout has also been simulated by a
group at Leiden University\cite{teitelbaum-archive} who have independently
pointed out that $^{139}$La NQR intensity provides valuable information
regarding
$\Gamma$.  They base their analysis on
$^{139}$La spin-lattice relaxation data in Nd codoped materials. 
Although we agree that
$^{139}1/T_1$ is a good place to initiate a study of the fluctuation
spectrum and Cu wipeout, we note that nuclear relaxation in the Nd
codoped materials is usually dominated by the large Nd moment, as in
NdBa$_2$Cu$_3$O$_7$\cite{itoh-Nd} and
Nd$_{2-x}$Ce$_x$CuO$_4$.\cite{zheng-Nd}  This also seems to be the case
in Nd doped La$_{2-x}$Sr$_x$CuO$_4$ as shown in Fig.~\ref{fig_1/T1T}. 
Furthermore, we find that $^{139}1/T_1$ in the Nd codoped materials is
roughly independent of hole doping.\cite{singer-unpublished}  Bearing this
in mind, it seems that their analysis would then predict identical wipeout
for different hole concentrations $x$, which is certaintly not the case
(see Fig.~\ref{fig_Cu_wipeout_fraction}).

\section{N\lowercase{d} and E\lowercase{u} codoped materials away from
$\bf \lowercase{x}\approx\frac{1}{8}$}
\label{Nd_Eu_phase_diagram}

\subsection{Construction of a unified phase diagram}

\begin{figure}
\begin{center}
\epsfig{figure=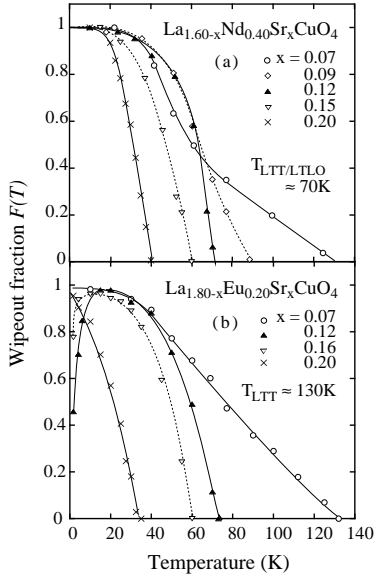,width=2.1in,angle=0}
\end{center}
\caption{ (a) The fraction of wiped out signal, $^{63}F(T)$, in
La$_{1.6-x}$Nd$_{0.4}$Sr$_{x}$CuO$_{4}$. The low temperature structural 
phase transition temperature $T_{LTT}$ 
 at $x\approx \frac{1}{8}$ is $T_{LTT} \approx 70$\,K. \cite{ichikawa} 
The lines are guides  for the eye and only data points below $T_{NQR}$ are
plotted. (b) is the same plot but  for
La$_{1.8-x}$Eu$_{0.2}$Sr$_{x}$CuO$_{4}$. At $x\approx \frac{1}{8}$,
$T_{LTT} \approx 130$\,K. \cite{kataev}}
\label{fig_Cu_wipeout_fraction}
\end{figure}

\begin{figure}
\begin{center}
\epsfig{figure=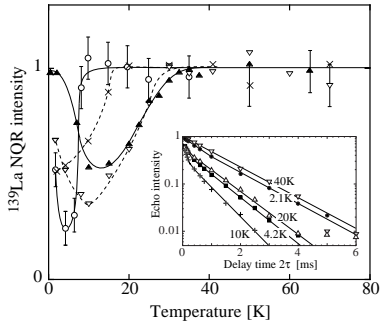,width=2.2in,angle=0}
\end{center}
\caption{Plot of the $^{139}$La NQR signal intensity
in La$_{1.8-x}$Eu$_{0.2}$Sr$_{x}$CuO$_{4}$ for the 
$|\pm\frac{5}{2}\rangle\leftrightarrow
|\pm\frac{7}{2}\rangle$ transition at a frequency $3 \nu_{Q} 
\approx$ 18\,MHz ($x=0.07, \circ; 0.12, \blacktriangle; 0.16,
\triangledown; 0.20, \times$). The intensity was measured for a fixed
pulse separation time of $\tau=40\,\mu$sec.  In the inset we show the
spin echo decay for La$_{1.68}$Eu$_{0.20}$Sr$_{0.12}$CuO$_4$ at different
temperatures.  All curves are guides for the eye.}
\label{fig_La_Eu_wipeout_fraction}
\end{figure}

\begin{figure}
\begin{center}
\epsfig{figure=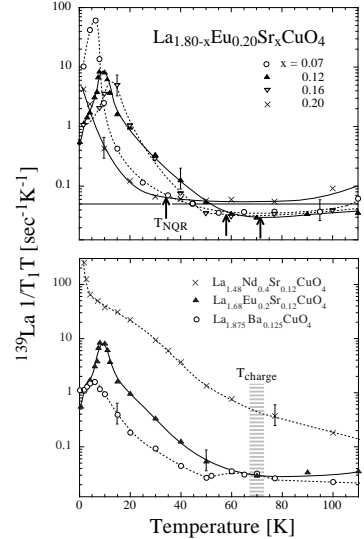,width=2.2in,angle=0}
\end{center}
\caption{(a) Plot of $^{139}1/T_{1}T$ of the
$|\pm\frac{5}{2}\rangle\leftrightarrow |\pm\frac{7}{2}\rangle$ transition
for La site in La$_{1.8-x}$Eu$_{0.2}$Sr$_{x}$CuO$_{4}$  where the symbols
used for each hole concentration $x$ is listed in  the figure. The
horizontal line marks the constant $^{139}1/T_{1}T \approx
0.05\,$sec$^{-1}$K$^{-1}$ value and the arrows mark $T_{NQR}$ for each
material $x=0.20$, 0.16 and 0.12 from left to right.  (b) The same
quantity, for several samples with hole concentration
$x\approx\frac{1}{8}$.  Note that the Nd codoped system has an enhanced
relaxation rate due to the Nd moments.  The bump in the data for the Ba
doped sample near 60\,K arises from the LTT structural transition.}
\label{fig_1/T1T}
\end{figure}

Now that we have reached a good quantitative understanding of the spin
stripe fluctuations and its temperature dependence  for the magic 
hole concentration of $x\approx\frac{1}{8}$, we seek a comprehensive 
understanding of stripe fluctuations across the entire phase diagram of  
La$_{1.6-x}$Nd$_{0.4}$Sr$_{x}$CuO$_{4}$ 
and La$_{1.8-x}$Eu$_{0.2}$Sr$_{x}$CuO$_{4}$.  The temperature 
dependence of the $^{63}$Cu NQR wipeout $^{63}F(T)$, $^{139}$La NQR
intensity, and  $^{139}1/T_{1}T$ are presented in
Figs.~\ref{fig_Cu_wipeout_fraction}, \ref{fig_La_Eu_wipeout_fraction},
and \ref{fig_1/T1T}, respectively.  Utilizing this information, as well as
available elastic neutron scattering and $\mu$SR measurements, we follow
the analysis of secton \ref{subsec_wipeout} to deduce the temperature
dependence of the spin fluctuation spectrum $\Gamma$ for hole
concentrations of $x=0.07, 0.12, 0.16$ and 0.20.  The results are depicted
in Fig.~\ref{fig_anchor_points} together with the simulated $^{63}$Cu
wipeout.

\begin{figure}
\begin{center}
\epsfig{figure=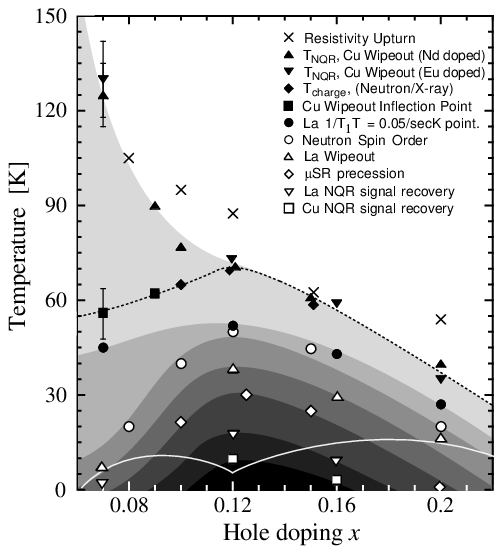,width=3.15in,angle=0}
\end{center}
\caption{Phase diagram for La$_{1.6-x}$Nd$_{0.4}$Sr$_{x}$CuO$_{4}$ 
and La$_{1.8-x}$Eu$_{0.2}$Sr$_{x}$CuO$_{4}$, showing the upturn in ab-plane
resistivity temperature $T_{u}$ found in
La$_{1.6-x}$Nd$_{0.4}$Sr$_{x}$CuO$_{4}$\cite{ichikawa} ($\times$), $^{63}$Cu
wipeout onset $T_{NQR}$ in
La$_{1.6-x}$Nd$_{0.4}$Sr$_{x}$CuO$_{4}$ ($\blacktriangle$) and 
La$_{1.8-x}$Eu$_{0.2}$Sr$_{x}$CuO$_{4}$ ($\blacktriangledown$), 
onset temperature $T_{charge}$ for short range charge order according 
to neutron\cite{tranquada2,ichikawa} 
and X-ray\cite{zimmermann,niemoller} in
La$_{1.6-x}$Nd$_{0.4}$Sr$_{x}$CuO$_{4}$
(\ding{117}), the copper wipeout inflection point ($\blacksquare$) in 
La$_{1.6-x}$Nd$_{0.4}$Sr$_{x}$CuO$_{4}$, 
the temperature where $^{139}1/T_{1}T = 0.05\,$sec$^{-1}$K$^{-1}$ in 
La$_{1.8-x}$Eu$_{0.2}$Sr$_{x}$CuO$_{4}$ ($\bullet$),
long range spin order $T_{spin}$ 
\cite{tranquada3,ichikawa} in La$_{1.6-x}$Nd$_{0.4}$Sr$_{x}$CuO$_{4}$
($\circ$), onset of La wipeout for La$_{1.8-x}$Eu$_{0.2}$Sr$_{x}$CuO$_{4}$
($\triangle$), onset of $\mu$SR coherent precession $T_{\mu
SR}$\cite{nachumi} for
La$_{1.6-x}$Nd$_{0.4}$Sr$_{x}$CuO$_{4}$ ($\diamond$), 
the onset of $^{139}$La recovery of signal in
La$_{1.8-x}$Eu$_{0.2}$Sr$_{x}$CuO$_{4}$ ($\triangledown$), and the onset
of $^{63,65}$Cu recovery of signal in
La$_{1.8-x}$Eu$_{0.2}$Sr$_{x}$CuO$_{4}$ ($\square$). The darker grey
tones indicate increasingly slow fluctuation timescales. We also show the
superconducting boundary as a white line and a dotted line that connects
$T_{charge}$ and the Cu wipeout inflection points as a guide for the eye.
} 
\label{fig_phase_diagram}
\end{figure}

Another way to illustrate the temperature dependence of the spin dynamics
of stripes is to plot contours of characteristic frequency scales on the
temperature ($T$) versus hole concentration ($x$) phase diagram.  In Fig.
\ref{fig_phase_diagram}, we collect data from various  experiments
and plot the temperature at which each experimental probe experiences an
anomaly.  The anomaly occurs at the temperature at which the spin
fluctuations of some parts of the CuO$_{2}$ plane slow  to the inherent
frequency of that probe.  Using the anomaly
temperatures as a boundary, we section off  areas of the phase diagram
into various shades of grey. 

The white  area at the top of the plot corresponds to the temperature
region with very fast spin fluctuations, where stripes can be considered
to be completely dynamic.  In other words, this is the region where the
so-called {\it
$\omega/T$ scaling} holds for spin dynamics, and there is a well-defined
spin fluctuation time scale whose distribution is minimal.  
(It should be noted that there is NMR evidence for some spatial
inhomoginiety even in a similar high temperature regime, as recently
reported by Haase, Slichter, and co-workers on the basis of some
ingenious NMR measurements and analysis,\cite{Haase-Slichter} and by a
report of the frequency dependence of $^{63}1/T_1$ by Fujiyama {\it
et\,al.}.\cite{fujiyama}

The darkest grey region shows where the spin fluctuations are so slow 
that the hyperfine field is nearly static even at the NQR times 
scales.  This region is obtained based on the appearance of hyperfine 
broadening in the recovered NQR signals of $^{63}$Cu.
Between the lightest and darkest grey areas, which correspond to the wipeout
and recovery of $^{63}$Cu NQR signal, there are five separate regions
depicted in Fig.~\ref{fig_phase_diagram}.  The boundary marked by
solid circles is set by the temperature at which $^{139}1/T_{1}T$ crosses the
value 0.05\,sec$^{-1}$K$^{-1}$  in La$_{1.8-x}$Eu$_{0.2}$Sr$_{x}$CuO$_{4}$
(see Fig.~\ref{fig_1/T1T}). $^{139}1/T_{1}T$ is somewhat
larger at high temperatures for $x=0.20$, so in this case we use the
temperature (25\,K) at which the enhancement of $^{139}1/T_{1}T$
begins.  The next boundary is set by $T_{spin}$, the onset of the elastic
neutron scattering from spin  stripes corresponding to a frequency scale
of $0.5-1.0$\,meV, based on the energy integration windows of those
experiments.   The next area marks the onset temperature of 
$^{139}$La NQR wipeout.  Open diamonds mark the onset temperature for muon
spin  rotation $T_{\mu SR}$ arising from static order on the
timescales of $\mu$SR measurements.\cite{nachumi}
At temperatures lower than 20\,K the NQR signal begins to recover for the 
$^{139}$La nuclei.  These temperatures are marked by open, downward pointing
triangles in Fig.~\ref{fig_phase_diagram}. Also shown in Fig.
\ref{fig_phase_diagram} are the onset temperatures $T_{charge}$ for short
range charge order as obtained from neutron scattering\cite{ichikawa} and
x-ray scattering,\cite{zimmermann,niemoller} the transition to the
superconducting phase in  La$_{1.6-x}$Nd$_{0.4}$Sr$_{x}$CuO$_{4}$ (a
white line), and a  dotted black line (as a guide for the eye) that
follows $T_{charge}$ and the Cu wipeout inflection points.  The
importance of the inflection points will be explained below.

\subsection{Results for $\bf x\gtrsim\frac{1}{8}$}

We have already discussed the sharp
onset of the slowing of spin fluctuations that occurs at $T_{charge}$
for $x=0.12$ which is caused by charge order.  Looking at the higher hole
concentrations, $x=0.16$ and 0.20, we see similar behavior in the
$^{63}$Cu NQR wipeout (Fig.~\ref{fig_Cu_wipeout_fraction}),  $^{139}$La
NQR wipeout (Fig.~\ref{fig_La_Eu_wipeout_fraction}), and in $^{139}1/T_1T$
(Fig.~\ref{fig_1/T1T}), although in each case the temperature scale is
shifted increasingly downwards for the samples farther from the magic hole
concentration $x\approx\frac{1}{8}$.  This naturally leads to a similar
distribution of $\Gamma$ that is shifted to lower temperature, as shown in
Fig.~\ref{fig_anchor_points}c and \ref{fig_anchor_points}d.  In the case
of $x=0.16$ we find that all of the anchoring points of the distribution
are shifted to lower temperatures by $\sim$10\,K while for $x=0.20$ the
shift is $\sim$20-30\,K.  The striking similarity of the shapes of
$\Gamma(T)$ for $0.12\leq x\leq0.20$ indicates that the fundamental
physics of these systems is very similar.  In each of these cases we
consider charge order as the cause of anomalous glassy slowing of spin
fluctuations.  With increasing hole concentration the tendency to charge
order diminishes, causing the shift of $\Gamma$ to lower temperatures for
larger $x$.  

The same strong similarities for the samples with
$x\gtrsim\frac{1}{8}$ can also be seen in the phase diagram of
Fig.~\ref{fig_phase_diagram}.  The lightest grey corresponds to the
region where some segments of the CuO$_2$ plane experience spin
fluctuations slower than $^{63}\Gamma^{wipeout}$, and $^{63}$Cu NQR
signal begins to wipe out.  For La$_{1.6-x}$Nd$_{0.4}$Sr$_{x}$CuO$_{4}$
with $0.12\lesssim   x\lesssim 0.15$, the boundary  agrees
well with the onset of  short-range charge order as observed by scattering
methods.\cite{tranquada2,zimmermann}  We note that our wipeout studies 
of La$_{1.6-x}$Nd$_{0.4}$Sr$_{x}$CuO$_{4}$ were conducted\cite{singer} 
without the knowledge of the more recent results from x-ray scattering by
Niem\"oller {\it et\,al.}\ for $x=0.15$,\cite{niemoller} and the good
agreement between our $T_{NQR}=60\pm$5\,K and  their $T_{charge}=55$\,K
strongly supports our  indentification of $T_{NQR}$ as the onset of the
glassy slowing of the  spin fluctuations triggered by slowing charge
dynamics for $0.12\lesssim  x\lesssim 0.15$.  Generally speaking, for 
$x\gtrsim\frac{1}{8}$, we find that all of the contours are nearly
parallel and slope downward with increasing $x$. Thus below
the onset temperature for charge order $T_{charge}$, the
slowing down is similarly rapid throughout this concentration range, but
with an energy scale that decreases away from the magic doping level
$x\approx\frac{1}{8}$ where the stripe fluctuations are most
robust.\cite{singer}   In fact, if we fit the slowing of spin
fluctuations to the renormalized classical form
$\exp(-2\pi\rho_s^{eff}/T)$ for different hole concentrations we find that
$2\pi\rho_s^{eff}$ decreases with increasing $x$ above
$x\approx\frac{1}{8}$ as shown in Fig.~\ref{fig_doping_of_rho}.

\subsection{Results for $\bf x\lesssim\frac{1}{8}$}
\label{subsec_underdoped}

Within the second region $x\lesssim\frac{1}{8}$, the slowing down is more
gradual and the contours in Fig.~\ref{fig_phase_diagram} are more spread
out.  At this point it is worth pointing out that although $T_{NQR}$
increases below $x \approx\frac{1}{8}$, $^{63}F(T)$ is more tailed as
shown in Fig.~\ref{fig_Cu_wipeout_fraction} for
La$_{1.6-x}$Nd$_{0.40}$Sr$_{x}$CuO$_{4}$ with $x=0.07$ and 0.09. It is
important to note that there is an inflection point in the temperature
dependence of  the wipeout fraction, $^{63}F(T)$, at approximately 55\,K
and 62\,K,  respectively, for these two concentrations.  In a recent
study on a La$_{1.50}$Nd$_{0.4}$Sr$_{0.10}$CuO$_{4}$ single crystal
provided by Ichikawa {\it et\,al.}, we confirmed\cite{singer-unpublished}
that this inflection point is nothing but the charge ordering temperature
$T_{charge}$ as determined by x-ray scattering.\cite{ichikawa}  The
coincidence of $T_{charge}$ and the LTLO (Low Temperature Less
Orthorhombic) structural phase transition temperature\cite{ichikawa} may
suggest  that static one dimensional charge order cannot develop without
the  tetragonal lattice symmetry that is compatible with the symmetry of 
charge stripes. The longer spatial coherence of charge order below
$T_{charge}$ accelerates wipeout at the inflection point. 
Even though static charge order with substantial spatial coherence  has
not developed between $T_{NQR}$ and $T_{charge}$, transport measurements
by Ichikawa  {\it et\,al.}\  show that  the onset of wipeout agrees well
with the temperature scale $T_{u}$  for {\it local charge order} as
deduced from a scaling analysis of  the enhancement of in-plane
resisitivity.\cite{ichikawa}  These results underscore the  importance of
the interplay between charge and spin degrees of freedom, even for
$x\lesssim\frac{1}{8}$.

\begin{figure}
\begin{center}
\epsfig{figure=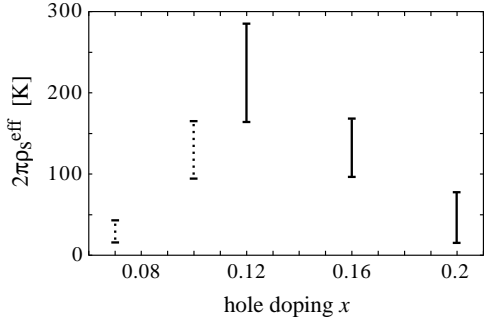,width=2.9in,angle=0}
\end{center}
\caption{Doping dependence of $2\pi\rho_s^{eff}$ as deduced from the data
presented above in Fig.~\ref{fig_phase_diagram} by fitting to the form
$\Gamma\propto\exp(-2\pi\rho_s^{eff}/T)$.  We note that
$\Gamma$ levels for low temperature and that region is neglected in the
fit, resulting in large uncertainty in the value of $2\pi\rho_s^{eff}$. 
For $x<\frac{1}{8}$ tailed wipeout indicates that the physics may differ
substantially from the region $x\gtrsim\frac{1}{8}$, so we use dashed
lines in that region.} 
\label{fig_doping_of_rho}
\end{figure}

It would thus appear that for $x\lesssim\frac{1}{8}$, there is a
precursive mechanism that wipes out the $^{63}$Cu resonance signal below
$T_{NQR}$ before collective short range charge order sets in at
$T_{charge}(<T_{NQR})$. The nature of this precursive wipeout cannot be 
collective since there is no sharp onset, and thus cannot be  explained
in the terms discussed in section \ref{section_slowing}. In earlier
publications,\cite{hunt,singer} we have suggested that this precursive
wipeout may be accounted for if we assume that individual holes begin to
localize with decreasing temeprature, thereby creating local moments. 
This is a situation similar to the one  that results in wipeout in
conventional dilute Kondo alloys  \cite{nagasawa} and spin glasses
\cite{MacLaughlin-Alloul,chen} On the other hand, in this scenario
in which the local moments are created by localization effects, the
precursive wipeout may be further  enhanced if one increases resistivity
by introducing disorder.  However, our recent experimental studies on Zn
doped La$_{2-x}$Sr$_{x}$CuO$_4$ show that the onset of wipeout
$T_{NQR}$ is a fairly well-defined temperature  scale that 
does not vary even when resistivity is enhanced and local moments
are introduced by Zn impurity doping.\cite{singer-unpublished} 
Furthermore, a recent inelastic neutron scattering study\cite{hiraka} on
a sample of La$_{1.93}$Sr$_{0.07}$CuO$_{4}$ showed a continual decrease
of the fluctuation energy scale with decreasing temperature, followed by
the freezing of spin stripes at low temperatures.  The presence of a
striped ground state indicates that it is unlikely that the glassy spin
fluctuations below $T_{NQR}$ have no spatially correlated stripe
signature.  We include data for $2\pi\rho_s^{eff}$ for the range
$x<\frac{1}{8}$ in Fig.~\ref{fig_doping_of_rho}, but we reiterate the
differences between the physics above and below $x\approx\frac{1}{8}$ and
use dotted lines for the underdoped samples.  We note that the trend
of $2\pi\rho_s^{eff}$ decreasing away from $x\approx\frac{1}{8}$ is
expected from the low temperature behavior seen in
Fig.~\ref{fig_phase_diagram}, indicating that the relevant energy scale
is indeed decreasing, even if an identical analysis is not entirely
applicable.

\section{CONCLUSIONS}

In this paper, we provide a comprehensive NQR picture of the spatial 
modulation and fluctuations of spin and charge density waves 
in the stripe phase of  
La$_{1.88-y}$(Nd,Eu)$_{y}$Sr$_{0.12}$CuO$_{4}$ and 
La$_{1.875}$Ba$_{0.125}$CuO$_{4}$.  From the analysis of the Zeeman 
perturbed $^{63}$Cu and $^{139}$La NQR spectra, we deduce
the spatial distribution of the hyperfine field at the time 
scales of NQR spin echo measurements.  We 
demonstrate that the maximum frozen Cu magnetic moment at 350\,mK
is relatively small, $\sim$$\,0.1-0.2\mu_{B}$, with a comparably large
distribution. The fact that $\mu$SR studies find a large and well defined
effective moment of $\sim$$\,0.3\,\mu_B$ at a timescale that is two orders
of magnitude faster than NQR, provides evidence that motional avergaing is
at least partially responsible for the supression of the effective moment
as observed in our NQR measurements. Given that
the mobility of holes in the CuO$_{2}$ planes at $T\ll T_{charge}$ is
comparable to the  metallic state,\cite{noda} perhaps it is reasonable
that the magnitude  of the frozen moment averaged over NQR time scales is
fairly small and  highly disordered, despite the fact that the spin-spin
correlation length  is known to reach
$\sim\,$200\,\AA\ below 30\,K at neutron scattering time
scales.\cite{tranquada4}  We also find that broadening of the quadrupole
coupling is necessary to fit the low temperature Zeeman perturbed NQR
linesape, indicating that there are spatial variations in the local charge
state.

From a simple physical argument based on 
comparison of $^{139}1/T_{1}$ and the 
renormalized classical scaling of the O(3) Non-Linear-$\sigma$ 
model, we show that the roughly exponential divergence of 
$^{139}1/T_{1}$ observed below $T_{charge}$ is consistent with the 
presence of low frequency spin fluctuations characterized by a reduced
effective spin stiffness $2\pi\rho_{s}^{eff}\approx200$\,K.  We
attribute the renormalization of the spin stiffness from
$2\pi\rho_s\sim10^3$\,K at higher temperatures to
$2\pi\rho_{s}^{eff}\sim10^2$\,K to slowing charge  dynamics that suppress
the Cu-Cu exchange interaction.   By analyzing the magnitude and
distribution of  the nuclear spin-lattice relaxation rate
$^{139}1/T_{1}$  based the Non-Linear-$\sigma$  model with
$2\pi\rho_{s}^{eff}\approx200$\,K, we deduce the spatial distribution of
$\Gamma$ (spin fluctuation frequency scale) of Cu below $T_{charge}$.  We
also analyze the $^{63}$Cu and $^{139}$La NQR wipeout fraction $F(T)$,
elastic neutron scattering,  and $\mu$SR data within the same framework,
and obtain a nearly  identical spectrum. Combined
with the fact that only a small change of a factor of $\sim\,$3 in the
anomalous slowing of $\Gamma$ is sufficient  to account for the observed
wipeout of $^{63}$Cu NQR, this provides a  natural explanation of why the
temperature dependence of $F(T)$ tracks  the charge order parameter
measured by scattering techniques for  $0.12<x<0.16$, which may be
interpreted as the volume fraction of patches of the sample in which the
spin fluctuations have started to slow exponentially.  However, we note
that our model is quite simple minded, relying entirely on magnetic
enhancement of $1/T_1$.\cite{detuned}  Fortunately, since the qualitative
change in the temperature dependence of spin fluctuations from roughly
$T$-linear to exponential is so drastic, the overall picture of the
temperature dependence of the spin fluctuation frequency scale $\Gamma$
shown in Fig.~\ref{fig_anchor_points} would not be substantially altered
by a more elaborate analysis.

By plotting contours of spin fluctuation time scales in 
the $T-x$ phase diagram, we demonstrate that similar glassy 
slowing of stripes takes place away from the magic hole concentration 
$x=\frac{1}{8}$.  However, we find a qualitative difference in the 
slowing between  above and below $x=\frac{1}{8}$ where the measured
incommensurability begins to saturate.\cite{yamada}  The origin  of the
difference is not well understood at this point.
For $x<\frac{1}{8}$, stronger tendency toward some sort of localized state
seems to precede spatially periodic charge ordering,\cite{noda,ichikawa}
probably inducing local spin glass like behavior with fairly high
temperature scales.\cite{hunt,singer,imai89}  However, we emphasize that
this spatially disordered spin glass like entity eventually freezes as
stripes, as demonstrated by neutron scattering.\cite{wakimoto}  Hence
$T_{NQR}$ may be considered as the onset of the glassy freezing of
stripes.  A recent scaling analysis of resistivity data by Ichikawa {\it
et\,al.}\cite{ichikawa} suggests that this glassy freezing is triggered
by slowing charge dynamics, in support of our earlier proposal for the
mechanism of wipeout and glassy freezing.\cite{singer}  We also believe
that the qualitative difference above and below $x\approx\frac{1}{8}$
bears  significant implications on the mechanism of superconductivity 
because  the behavior of La$_{2-x}$Sr$_{x}$CuO$_{4}$ is very similar to  
La$_{1.88-y}$(Nd,Eu)$_{y}$Sr$_{0.12}$CuO$_{4}$ at and below 
$x=\frac{1}{8}$, while the glassy freezing of stripes is not seen by NQR
in superconducting La$_{2-x}$Sr$_{x}$CuO$_{4}$ for
$x>\frac{1}{8}$.   

Finally, we emphasize that the so-called stripe phase never looks striped
at the slow time scales of NQR measurements even for $x\approx\frac{1}{8}$
at temperatures as low as 350\,mK.  It is certaintly possible to argue
that the segments of the CuO$_2$ plane in which the $^{63}$Cu NQR signal
is wiped out are very similar to a conventional spin glass, and at slow
time scales those segments are phase separated from other patches of the
CuO$_2$ plane where the Cu spins still behave as they did above
$T_{charge}$.  Whether such segments are fixed in space is not clear.  The
essence of the stripe physics is that these disordered patches of the
CuO$_2$ planes with {\it slow} spin and charge dynamics appear striped if
viewed by an experimental probe with a timescale that is faster than the
remnant fluctuations.

\section{ACKNOWLEDGMENTS}
We thank many colleagues for sharing their wisdom with us, 
in particular J.M. Tranquada, P.A. Lee, S. Uchida, H. Eisaki, G.S. 
Boebinger, Y. Ando, Y.S. Lee, M.A. Kastner, R.J. Birgeneau, N.J. Curro,
P.C. Hammel, J. Zaanen,  H. Brom, J. Haase, and C.P. Slichter among
others.    This work was financially supported by NSF DMR-9971264  
and NSF DMR 98-08941 through the MRSEC program.


\begin{references}


\bibitem{theory} For recent theoretical discussions, see 
A.H. Castro Neto and D. Hone, Phys. Rev. Lett {\bf 76}, 2165 (1996);
C.N.A. van Duin and J. Zaanen, Phys. Rev. Lett {\bf 80}, 1513 (1998);
S.R. White and D.J. Scalapino, Phys. Rev. Lett. {\bf 80}, 1272 (1998);
S.A. Kivelson, E. Fradkin,  V.J. Emery, Nature {\bf 393}, 550 (1998);
M. Vojta and S. Sachdev, Phys. Rev. Lett {\bf 83}, 3916 (1999);
L.P. Pryadko, S.A. Kivelson, V.J. Emery, Y.B. Bazaliy, and E.A. Demler,
Phys. Rev. B {\bf 60}, 7541 (1999).  For earlier theoretical works see J.
Zannen and O. Gunnarsson, Phys. Rev. B {\bf 40}, 7391 (1989); D. Poiblanc
and T.M.  Rice, Phys. Rev. B {\bf 39}, 9749 (1989); K. Machida, Physica C
{\bf 158}, 192 (1989);  V.J. Emery, S.A. Kivelson, and H.-Q. Lin, Phys.
Rev. Lett {\bf 64}, 475 (1990); and C. Castellani, C. Di Castro, and M.
Grilli, Phys. Rev. Lett {\bf 75}, 4650 (1995).


\bibitem{tranquada1} J.M. Tranquada, B.J. Sternlieb, J.D. Axe, Y. 
                   Nakamura, and S. Uchida, Nature {\bf 375}, 561  (1995). 
\bibitem{suzuki1998} T. Suzuki, T. Goto,\,K. Chiba, T. Shinoda, T. Fukase, 
                   H. Kimura, K. Yamada, M. Ohashi, and Y. Yamaguchi,
                   Phys. Rev. B {\bf 57}, R3229  (1998).
\bibitem{kimura} H.\,Kimura, H. Hirota, H. Matsushita,\,K. Yamada, Y. Endoh,
                   S.-H. Lee, C.F. Majkrzak, R. Erwin, G. Shirane, M. Greven,
                   Y.S. Lee, M.A.\,Kastner, and R.J. Birgeneau, Phys. Rev. B
                   {\bf 59}, 6517 (1999).
\bibitem{wakimoto} S. Wakimoto, G. Shirane, Y. Endoh, K. Hirota, S. Ueka,
                   K. Yamada, R.J. Birgeneau, M.A.\,Kastner, Y.S. Lee,
                   P.M. Gehring, and S.-H.  Lee, cond-mat/9902201.
\bibitem{lee} Y.S. Lee, R.J. Birgeneau, M.A.\,Kastner, Y. Endoh, S. 
                   Wakimoto,\,K. Yamada, R.W. Erwin, S.-H, Lee, G. Shirane, 
                   cond-mat/9902157.  
\bibitem{tranquada2} J.M. Tranquada, J. D. Axe, N. Ichikawa, Y. 
                   Nakamura, S. Uchida, and B. Nachumi, Phys. Rev. B
                   {\bf 54}, 7489 (1996).
\bibitem{tranquada3} J.M. Tranquada, J.D. Axe, N. Ichikawa, A.R. 
                   Moodenbaugh, Y. Nakamura, and S. Uchida, Phys. Rev. 
                   Lett. {\bf 78}, 338 (1997).
\bibitem{tranquada4} J.M. Tranquada, N. Ichikawa, and S. Uchida,
                   Phys. Rev. B {\bf 59}, 14712 (1999).
\bibitem{hunt} A.W. Hunt, P.M. Singer,\,K.R. Thurber, and T. Imai, 
                   Phys. Rev. Lett. {\bf 82}, 4300 (1999).
\bibitem{singer} P.M. Singer, A.W. Hunt, A.F. Cederstr\"om, and T. Imai, Phys.
                  Rev. B {\bf 60}, 15354 (1999).
\bibitem{curro} N.J. Curro, P.C. Hammel, B.J. Suh, M. H\"ucker, B. B\'ucher, U.
                   Ammerahl, and A. Revcolevscji, Phys. Rev. Lett.     
                   {\bf 85}, 642 (2000).
\bibitem{teitelbaum} G.B. Teitel'baum, B. B\"uchner, and H. de 
                   Gronckel, Phys. Rev. Lett. {\bf 84}, 2949 (2000).
\bibitem{julien}M.-H. Julien, A. Campana, A. Rigamonti, P. Carretta, F.
                    Borsa, P. Kuhns, A.P. Reyes, W.G. Moulton, M.
                    Horvati\'c, C. Berthier, A. Vietkin, and A.
                    Revcolevschi, cond-matt/0010362.
\bibitem{rameev} B. Rameev, E.\,Kukovitskii, V.\,Kataev, and G.
                   Teitel'baum, Physica C {\bf 246}, 309 (1995).    
\bibitem{kataev} V. Kataev, B. Rameev, A. Validov, B. B\"uchner, 
                   M. H\"ucher, and R. Borowski, Phys. Rev. B {\bf 58},
                   R11876 (1998).
\bibitem{zimmermann} M.v. Zimmermann, A. Vigliante, T. Niem\"oller, N. 
                   Ichikawa, T. Frello, J. Madsen, P. Wochner, S. Uchida, 
                   N.H. Andersen, J.M. Tranquada, D. Gibbs, and J.R. 
                   Schneider, Europhys. Lett. {\bf  41}, 629 (1998).
\bibitem{niemoller} T. Niem\"oller, H. H\"unnefeld, J.R. Schneider, N. 
                   Ichikawa, S. Uchida, T. Frello, N.H. Andersen, and 
                   J.M. Tranquada, cond-mat/9904383.  
\bibitem{ando-boebinger} Y. Ando, G.S. Boebinger, A. Passner, T. Kimura
                   and K. Kishio, Phys. Rev. Lett. {\bf 75}, 4662 (1995). 
\bibitem{boebinger}G.S. Boebinger, Yoichi Ando, A. Passner, T.\,Kimura,
                   M. Okuya, J. Shimoyama,\,K.\,Kishio,\,K. Tamasaku, N.
                   Ichikawa, and S. Uchida, Phys. Rev. Lett. {\bf  77},
                   5417 (1996).
\bibitem{noda} T. Noda, H. Eisaki, and S. Uchida, Science {\bf 286}, 265
                   (1999).
\bibitem{ichikawa} N. Ichikawa, S. Uchida, J.M. Tranquada, 
                   T. Niemoeller, P.M. Gehring, 
                   S.-H. Lee, J.R. Schneider, Phys. Rev. Lett {\bf 85},
                   1738 (2000).
\bibitem{uchida} S. Uchida, N. Ichikawa, T. Noda, and H. Eisaki, to be
                   published in Vol. 125 
                   Springer Series in Solid State Sciences.
\bibitem{ichikawa-thesis} N. Ichikawa, S. Uchida, J.M. Tranquada, T. Niem\"oller, 
                   P.M. Gehring, S.-H. Lee, and J.R. Schneider in preparation; 
                   N. Ichikawa, thesis, University of Tokyo (1999).
\bibitem{Fujimori} A. Ino, T. Mizokawa, K. Kobayashi, A. Fujimori, T. 
                   Sasagawa, T. Kimura, K. Kishio, K. Tamasaku, H. Eisaki,
                   and S. Uchida, Phys. Rev. Lett. {\bf 81}, 2124 (1998).
\bibitem{zhou} X.J. Zhou, P. Bogdanov, S.A. Keller, T. Noda, H. Eisaki, S.
                   Uchida, Z. Hussain, and Z.-X. Shen, Science {\bf 286},
                   268 (1999).
\bibitem{luke} G.M. Luke, L.P. Le, B.J. Strenlieb, W.D. Wu, Y.J. 
                   Uemura, J.H. Brewer, T.M. Riseman, S. Ishibashi, 
                   Physica C {\bf 185-189}, 1175 (1991).
\bibitem{kumagai}K.-I Kumagi, K. Kawano, I. Watanabe, K. Nishiyama,
                   and K.Nagamine, Hyperfine Int. {\bf 86}, 473 (1994).   
\bibitem{nachumi} B. Nachumi, Y. Fudamoto, A. Keren, K.M. Kojima, M. 
                   Larkin, G.M. Luke, J. Merrin, O. Tchernyshyov, Y.J.
                   Uemura, N.  Ichikawa, M. Goto, H. Takagi, S. Uchida,
                   M.K. Crawford, E.M.  McCarron, D.E. MacLaughlin, and
                   R.H. Heffner, Phys. Rev. B {\bf 58},  8760 (1998).
\bibitem{niedermayer} Ch. Niedermayer, C. Bernhard, T. Blasius, A. 
                   Golnik, A Moodenbaugh, and J.I. Budnick, Phys. Rev. 
                   Lett.   {\bf 80}, 3843 (1998).
\bibitem{kojima}K.M. Kojima, H. Eisaki, S. Uchida, Y. Fudamoto, I.M.
                   Gat,  A. Kinkhabwala, M.I. Larkin, G.M. Luke, Y.J.
                   Uemura, Physica B {\bf  289-290}, 343 (2000).
\bibitem{savici} A.T. Savici, Y. Fudamoto, I.M. Gat, M.I. Larkin, Y.J. 
                   Uemura, G.M. Luke, K.M. Kojima, Y.S. Lee, M.A. Kastner,
                   and R.J.  Birgeneau, Physica B {\bf 289-290}, 338
                   (2000).
\bibitem{imai89}T. Imai, K. Yoshimura, T. Uemura, H. Yasuoka,
                  and K. Kosuge, J. Phys. Soc. Jpn. {\bf 59}, 3846 
                  (1990).
\bibitem{yoshimura92}K. Yoshimura, T. Uemura, M. Kato, T.
                   Shibata, K. Kosuge,  T. Imai and H. Yasuoka
                   ``{\it The Physics and Chemistry of Oxide
                   Superconductors}," editors  Y. Iye and  H. Yasuoka,
                   Springer-Verlag Berlin, 405-407 (1992).
\bibitem{tou} H. Tou, M. Matsumura, and H. Yamagata, 
                   J. Phys. Soc. Jpn. {\bf 61}, 1477 (1992).
\bibitem{cho}J.H. Cho, F. Borsa, D.C. Johnston, and D.R. Torgenson, 
                   Phys. Rev. {\bf B 46}, 3179 (1992).  
\bibitem{chou}F.C. Chou, F. Borsa, J.H. Cho, D.C. Johnston, A. 
                   Lascialfari, D.R. Torgeson, and J. Ziolo, Phys. Rev.
                   Lett. {\bf 71},  2323 (1993).
\bibitem{goto} T. Goto, S.\,Kazama,\,K. Miyagawa, and T. Fukase, 
                   J. Phys. Soc. Jpn. {\bf 63}, 3494 (1994).
\bibitem{ohsugi1}S. Ohsugi, Y. Kitaoka, H. Yamanaka, K. Ishida, and K.
                   Asayama, J. Phys. Soc. Jpn. {\bf 63}, 2057 (1994).
\bibitem{ohsugi2}S. Ohsugi, J. Phys. Soc. Jpn. {\bf 64}, 3656 (1995).
\bibitem{abragam}A. Abragam, {\it Principles of Nuclear Magnetism}
                   (Oxford University Press, 1978).
\bibitem{slichter}  C.P. Slichter, {\it Principles of Magnetic 
                   Resonance} (Springer-Verlag, New York, 1989), 3rd ed.     
\bibitem{keimer} B. Keimer, N. Belk, R.J. Birgeneau, A. Cassanho, C.Y. 
                   Chen, M. Greven, M.A. Kastner, A. Aharony, Y. Endoh,
                   R.W. Erwin, and G. Shirane, Phys. Rev. B {\bf
                   46}, 14034 (1992). 
\bibitem{sternlieb} B.J. Sternlieb, M. Sato, S. Shamoto, G. Shirane, 
                   and J.M. Tranquada, Phys. Rev. B. {\bf 47}, 5320
                   (1993).                   
\bibitem{aeppli} G. Aeppli, T.E. Mason, S.M. Hayden, H.A. Mook and
                   J. Kulda, Science {\bf 278}, 1432 (1997).
\bibitem{breuer} M. Breuer, B. B\"uchner, R. M\"uller, M. Cramm, 
                   O. Maldonado, A. Freimuth, B. Roden, R. Borowski, B. 
                   Heymer, and D.  Wohlleben, Physica C {\bf 208}, 217 (1993).
\bibitem{buchner94}B. B\"uchner, M. Breuer, M. Cramm, A.
                   Freimuth, H. Micklitz, W. Schlabitz, and A.P. Kampf,
                   J. Low. Temp. Phys. {\bf 95}, 285 (1994).   
\bibitem{yasuoka-old} H. Yasuoka, J. Phys. Soc. Jpn. {\bf 19}, 1182 
                   (1964). 
\bibitem{benedek} P. Heller and G. Benedek, Phys. Rev. Lett. {\bf 8}, 428
                   (1962).
\bibitem{tsuda}T. Tsuda, T. Shimizu, H. Yasuoka, K. Kishio, and K.
                   Kitazawa, J. Phys. Soc. Jpn. {\bf 57}, 2908
                   (1988).                   
\bibitem{yasuoka} H. Yasuoka, T. Shimizu, Y. Ueda, K. Kosuge, 
                   J. Phys. Soc. Jpn. {\bf 57}, 2659
                   (1988).  
\bibitem{maclaughlin} D.E. MacLaughlin, J.P. Vithayathil, H.B. Brom, 
                   J.C.J.M. de Rooy, P.C. Hammel, P.C. Canfield, A.P. 
                   Reyes, Z. Fisk, J.D. Thomson, and S-W. Cheong, Phys. 
                   Rev. Lett. {\bf 72}, 760 (1994).
\bibitem{lombardi} A. Lombardi, M. Mali, J. Roos, and D. Brinkman, Phys. 
                   Rev. {\bf B54}, 93 (1996).
\bibitem{mila-rice}F. Mila, and T.M. Rice, Physica C {\bf 157}, 561
                   (1988). 
\bibitem{vaknin} D. Vaknin, S.K. Sinha, D.E. Moncton,
                   D.C. Johnston, J.M. Newsam, C.R. Safinya, and
                   H.E. King Jr., Phys. Rev. Lett. {\bf 58}, 2802 (1987).
\bibitem{imai93} T. Imai, C.P. Slichter, K. Yoshimura, and K. Kosuge,
                  Phys. Rev. Lett. {\bf 70}, 1002 (1993).
\bibitem{uemura} Y.J. Uemura, W.J. Kossler, X.H. Yu, J.R. Kempton,  
                   H.E. Shone, D. Opie, C.E. Stronach, D.C. Johnson, 
                   M.S. Alverez and D.P. Goshorn, Phys. Rev. Lett., 
                   {\bf 59}, 1045 (1987);
                   S. Uchida, Physics C {\bf 185-189}, 1175 (1991).
\bibitem{yoshimura-imai}\,K. Yoshimura, T. Imai, T. Shimizu, Y.
                   Ueda, K. Kosuge, H. Yasuoka, J. Phys. Soc. Jpn.  
                   {\bf 58}, 3057
                   (1989).
\bibitem{yoshimura}\,K. Yoshimura, T. Uemura, M. Kato, K. Kosuge, 
                   T. Imai, and H. Yasuoka, Hyperfine Int. {\bf 
                   79}, 876 (1993). 
\bibitem{imai-unpublished}Unpublished $^{63}$Cu NQR spectra observed for
                   $x=0.02$, 0.04, 0.075, and 0.15 in 
                   La$_{2-x}$Sr$_{x}$CuO$_{4}$ measured up to 800\,K for  
                   Imai {\it et\,al.}.\cite{imai93}	
\bibitem{kambe}S. Kambe, H. Yasuoka, H. Takagi, S. Uchida, and Y. Tokura,
                    J. Phys. Soc. Jpn. {\bf 60}, 400 (1991).
\bibitem{nishihara}H. Nishihara, H. Yasuoka, T Shimizu, T. Tsuda, T.
                   Imai, S. Sasaki, S.\,Kanbe,\,K.\,Kishio,\,K.\,Kitazawa,
                   and\,K. Fueki, J. Phys. Soc. Jpn. {\bf 56}, 4559
                   (1987).  	                     
\bibitem{kitaoka} Y. Kitaoka, S. Hiramatsu, K. Ishida, T. Kohara,
                   and K. Asayama, J. Phys. Soc. Jpn. {\bf 56}, 3024
                   (1987).                   
\bibitem{muha} G.M. Muha, Journal of Magnetic Resonance {\bf 53}, 85 (1983). 
\bibitem{kubo}R. Kubo and T. Toyabe, in the {\it Proceedings of Colloque 
                   Ampere XIV}, North Holland Publ. Co., 810 (1967).
\bibitem{uemura80} Y.J. Uemura, T. Yamazaki, R.S. Hayano, R. Nakai, and
                   C.Y. Huang, Phys. Rev. Lett. {\bf 45}, 583 (1980).
\bibitem{teitelbaum2} E. Vavilova, E.\,Kukovitskij, and G.B. Teitel'baum, 
                   Physica B {\bf 280}, 205 (2000).
\bibitem{teitelbaum-archive}G.B. Teitel'baum, I.M. Abu-Shiekah, O. 
                    Bakharev, H.B. Brom, and J. Zaanen, cond-mat/0007057.
\bibitem{suh-rapid}B.J. Suh, P.C. Hammel, M. H\"ucker, B. B\"uchner, U.
                    Ammerahl, and A. Revcolevschi, Phys. Rev. B {\bf 61},
                    R9265 (2000).
\bibitem{yamada}\,K. Yamada, C.H. Lee, K. Kurahashi, J. Wada, 
                    S. Wakimoto, S. Ueki, H. Kimura, Y. Endoh, S.
                    Hosoya, G. Shirane, R.J. Birgeneau, M. Greven,
                    M.A. Kastner, Phys. Rev. B {\bf 57}, 6165 (1998).


\bibitem{note-log-scales}  
Note that the contours plotted in Figs.~\ref{fig_anchor_points} and
\ref{fig_distribution_slices}b represent the function $fD(T,f)$ where the
extra factor of $f$ arises from
$\int D(T,f)df=\int fD(T,f)d(\log f)$.  Plotting $fD(T,f)$
allows direct integration in $d(\log f)$.
Similarly, in Figs.~\ref{fig_T1_recovery} and
\ref{fig_distribution_slices}a the distribution of
$T_1$ is plotted with a multiplicative factor of $T_1$ so that it can be
easily integrated by eye.  The distribution is actually a Gaussian in
$\log T_1$ divided by $T_1$. 


\bibitem{thurber-3leg}K.R. Thurber, T. Imai, T. Saito, M. Azuma, M. 
                  Takano, and F.C. Chou, Phys. Rev. Lett. {\bf 84},
                  558 (2000).
\bibitem{imai93-2}T. Imai, C.P. Slichter, K. Yoshimura, M. Katoh, and K.
                  Kosuge, Phys. Rev. Lett. {\bf 71}, 1254 (1993).
\bibitem{hayden91}S.M. Hayden, G. Aeppli, R. Osborn, A.D. Taylor, T.G.
                  Perring, S.-W. Cheong and Z. Fisk, Phys. Rev. Lett   
                  {\bf 67}, 3622 (1991).
\bibitem{singh-raman}R.R.P. Singh, P.A. Fleury, K.B. Lyons, and P.E.
                  Sulewski, Phys. Rev. Lett. {\bf 62}, 2736 (1989).
\bibitem{kim}Y.J. Kim, R.J. Birgeneau, M.A. Kastner, Y.S. Lee, Y. Endoh,
                    G. Shirane and K. Yamada, Phys. Rev. B {\bf 60}, 3294
                    (1999).
\bibitem{suh}B.J. Suh, P.C. Hammel, Y. Yoshinari, J.D. Thompson, J.L.
                    Sarrao, and Z. Fisk, Phys. Rev. Lett. {\bf 81}, 2791
                    (1998).
\bibitem{moriya56}T. Moriya, Prog. Theor. Phys {\bf 16}, 23 (1956).
\bibitem{charravarty-orbach}S. Chakravarty and R. Orbach, Phys. Rev.
                    Lett. {\bf 64}, 224 (1990).
\bibitem{moriya-gaussian} T. Moriya, Prog. Theor. Physics (Kyoto)
                    {\bf  16}, 641 (1956).


\bibitem{T-NQR} In our earlier publications\cite{hunt,singer} we did 
not distinguish between $T_{NQR}$ and $T_{charge}$.  However, recent 
scattering measurements by Ichikawa {\it et\,al.}\cite{ichikawa} showed 
that the onset of $^{63}$Cu NQR wipeout preceeds the onset of charge
order with short range spatial coherence at $T_{charge}$ for
$x < 0.12$.  Accordingly, we identify  the onset of wipeout as $T_{NQR}$
hereafter.  This is the same  notation which we originary used in the
initial version of Hunt {\it et\,al.}\cite{hunt} (see cond-mat/9902348,
version 1).  Upon a referee's  urgence, who pointed out that the usage of
too many newly defined physical parameters was confusing, we
eliminated many new symbols including $T_{NQR}$, assuming that
$T_{NQR}=T_{charge}$ for all concentrations.


\bibitem{abu} I.M. Abu-Shiekah, O.O. Bernal, A.A. Menovsky, H. Brom, 
                   and J. Zaanen, Phys. Rev. Lett. {\bf 83}, 3309
                   (1999). 


\bibitem{Ni-charge} The $^{139}$La NMR wipeout observed by Abu-Shiekah 
{\it et\,al.}\cite{abu} below 200\,K in La$_{2}$NiO$_{4+\delta}$ is not 
necessarily caused directly by charge fluctuations by themselves, though
the  temperature dependence of the wipeout fraction tracks the charge 
order parameter measured by neutron scattering quite well.  Because of the 
strong hybridization of the $^{139}$La atomic orbitals to Ni spin 
with 3d$^{8}$ configuration, the magnetic hyperfine interaction between Ni and La is 
sizable.  This means that if Ni d-spin fluctuations slow down below 
charge ordering, $^{139}$La NMR signal can be wiped out by the 
{\it indirect}, slowing spin mechanism caused by charge order.  This 
scenario is particularly plausible for La$_{2}$NiO$_{4+\delta}$ because 
application of high magnetic field to detect NMR is known to cause 
extra slowing down of spins.\cite{Wochner}     


\bibitem{tranquada7}J.M. Tranquada, P. Wochner, A.R. Moodenbaugh, and 
                   D.J. Buttrey, Phys. Rev. B {\bf 55}, R6113 (1997).
\bibitem{Wochner}P. Wochner, J.M. Tranquada, D.J. Buttrey, and V. 
                   Sachan, Phys. Rev. B {\bf 57}, 1066 (1998).
\bibitem{butaud} P. Butaud, P. S\'egaransan, C. Berthier, J. Dumas, 
                   and C. Shlenker, Phys. Rev. Lett. 
                   {\bf 55}, 253 (1985).
\bibitem{ross} J.H. Ross Jr., Z. Wang, and C.P. Slichter, Phys. Rev.
                   Lett. {\bf 56}, 663 (1986). 
\bibitem{nomura}\,K. Nomura, T. Sambongi,\,K.\,Kume, and M. Sato, Physica 
                   B {\bf 143}, 117 (1986). 
\bibitem{MacLaughlin-Alloul} D.E. MacLaughling and H. Alloul, Phys. Rev. 
                  Lett. {\bf 36}, 1158 (1976).
\bibitem{nagasawa} H. Nagasawa and W.A. Steyert, J. Phys. Soc. Jpn. 
                    {\bf 28}, 1171 (1970).
\bibitem{borsa} F. Borsa, M. Corti, T. Goto, A. Rigamonti, D.C. 
                  Johnstom, and F.C. Chou, Phys. Rev. B {\bf 45}, 5756
                  (1992).
\bibitem{hayden}S.M. Hayden, G. Aeppli, H.A. Mook, T.G. Perring, 
                  T.E. Mason, S.-W. Cheong, and Z. Fisk, Phys. Rev.
                  Lett. {\bf 76}, 1344  (1996).
\bibitem{johnston}D.C. Johnston, Phys. Rev. Lett. {\bf 62}, 957 (1989).
\bibitem{hwang}H.Y. Hwang, B. Batlogg, H. Takagi, H.L. Kao, R.J. Cava,
                  J.J. Krajewski, and W.F. Peck, Jr., Phys. Rev. Lett.
                  {\bf 72}, 2636 (1994).
\bibitem{chen} M.C. Chen and C.P. Slichter, Phys. Rev. {\bf B 27}, 
                  278 (1983). 
\bibitem{murani} A. P. Murani, J. Magn. Matter. {\bf 22}, 271 
                    (1981).
\bibitem{sera}M. Sera, Y. Ando, S. Kondoh, K. Fukuda, M. Sato, I.
                    Watanabe, S. Nakashima, and K. Kumagai, Solid State
                    Comm. {\bf 69}, 851 (1989).
\bibitem{ikegawa}S. Ikegawa, T. Yamashita, T. Sakurai, R. Itti, H.
                    Yamauchi, and S. Tanaka, Phys. Rev. B {\bf 43}, 2885 
                    (1991).
\bibitem{imai-houston} T. Imai, Invited talk presented at the
                   International Conference on Materials and Mechanisms of
                   Superconductivity and High Temperature Superconductors
                   VI, Houston, Texas, February 20-25, 2000.
\bibitem{itoh-Nd}G.Q. Zheng, Y. Kitaoka, Y. Oda, and K. Asayama, J. Phys.
                    Soc. Jpn. {\bf 58}, 1910 (1989).
\bibitem{zheng-Nd}M. Itoh, K. Karashima, M. Kyogaku, and I. Aoki, Physica
                    C {\bf 160}, 177 (1989).
\bibitem{singer-unpublished}P.M. Singer, A.W. Hunt, T. Imai, N. Ichikawa,
                    and S. Uchida, unpublished.
\bibitem{Haase-Slichter} J. Haase, R. Stern, C.T. Milling, C.P. Slichter, 
                    and D.G. Hinks, preprint.
\bibitem{fujiyama}S. Fujiyama, Y. Itoh, H. Yasuoka, and Y. Ueda, J. Pys.
                    Soc. Jpn. {\bf 66}, 2864 (1997).
\bibitem{hiraka}H. Hiraka, Y. Endoh, M. Fujita, K. Yamada, Y.S. Lee, A.
                    Ivanov, and J. Kulda, ``Novel Quantum Phenomena in
                    Transition Metal Oxides," Ministry of Education,
                    Science, Sports and Culture of Japan, Dec.\ 1999,
                    P3-35, p.~194.


\bibitem{detuned} We recall the presence of enhanced Lorentzian
$^{63}1/T_2$ near $T_{NQR}$ due to magnetic
fluctuations.\cite{hunt,singer}  We also recall the possibility of
contributions from {\it direct} charge contributions to
relaxation and wipeout.\cite{hunt,singer}


%
%
%
%
%
%
%
%
%
%


\end{references}
\end{document}